\documentclass[aps,a4paper,showkeys,nofootinbib,longbibliography,notitlepage,twocolumn,superscriptaddress]{revtex4-2}
\usepackage[utf8]{inputenc}
\usepackage[T1]{fontenc}
\usepackage{lmodern}
\usepackage{microtype}
\usepackage{filecontents}

\usepackage{natbib}

\usepackage[usenames,svgnames]{xcolor}
\usepackage{natbib}
\usepackage[hyperindex,breaklinks]{hyperref}
\hypersetup{
     colorlinks=true,       		
     linkcolor=Navy,          	
     citecolor=Navy,            
     filecolor=Navy,      		
     urlcolor=Navy,           	
    runcolor=cyan,
 }
\newcommand{\bra}[1]{\langle #1 |}
\newcommand{\ket}[1]{| #1 \rangle}

\newcommand{\tr}[0]{ {\rm tr}}

\newcommand{\half}[1]{{ \rm h}}

\newcommand{\SR}{{\rm SR}}

\newcommand{\diam}{{\rm diam}}

\usepackage{graphicx}
\usepackage{amsfonts}
\usepackage{amssymb}
\usepackage{amsmath}
\usepackage{enumitem}
\usepackage{booktabs}
\usepackage{placeins}

\newcounter{mainresult}
\newcommand{\mainresult}[2]{%
  \par\addvspace{0.75\baselineskip}%
  \refstepcounter{mainresult}\label{#1}%
  \noindent\textbf{Result~\themainresult. #2.}\par\nobreak
  \noindent\ignorespaces
}

\def\beq{\begin{align}}
\def\eeq{\end{align}}
\def\nbeq{\begin{equation*}}
\def\neeq{\end{equation*}}
\def\<{\langle}
\def\>{\rangle}

\def\tr{{\rm tr}}

\usepackage{mathtools}
\def\multiset#1#2{\ensuremath{\left(\kern-.3em\left(\genfrac{}{}{0pt}{}{#1}{#2}\right)\kern-.3em\right)}}

\usepackage{mathrsfs}
\newtheorem{appthm}{Theorem}[section]
\newtheorem{applemma}[appthm]{Lemma}
\newtheorem{appcor}[appthm]{Corollary}
\newtheorem{appprop}[appthm]{Proposition}
\newenvironment{appproof}[1][Proof]{%
  \par\noindent\textit{#1.}\ \ignorespaces
}{%
  \unskip\nobreak\hfill\ensuremath{\square}\par\medskip
}

\begin{document}

\title{The R\'enyi area law conjecture is false below \texorpdfstring{$\alpha=1$}{alpha = 1} in high dimensions}

\author{Tomotaka Kuwahara}
\affiliation{Analytical Quantum Complexity RIKEN Hakubi Research Team, RIKEN Center for Quantum Computing (RQC), Wako, Saitama 351-0198, Japan}
\affiliation{RIKEN Pioneering Research Institute (PRI), Wako, Saitama 351-0198, Japan}

\begin{abstract}
Area laws express the expectation that entanglement in gapped quantum ground states scales with a boundary rather than a volume. In one dimension this holds for every R\'enyi entropy, and the R\'enyi area-law conjecture asserts the same in higher dimensions for each fixed $0<\alpha<1$. Here we disprove this conjecture in every spatial dimension two and higher. We construct finite-range Hamiltonians with fixed local dimension, bounded interaction strength, unique ground states and a uniform spectral gap, whose ground states exhibit volume-law R\'enyi entanglement while obeying a von Neumann area law and remaining connected to product states through uniformly gapped local paths. The mechanism is error-space complexity: locally identifiable components of arbitrarily small weight can encode states with high entanglement complexity. We also establish the worst-case rank-error trade-off for approximate ground-state projectors, a key tool in one-dimensional area-law proofs. At fixed gap, the optimal Schmidt rank at projection error $\epsilon$ across a planar cut of cross-section $A$ scales as $\exp[\tilde{\Theta}((A\log\epsilon^{-1})^{2/3})]$ when $\log\epsilon^{-1}\gtrsim A$. Thus $\alpha=1$ is the sharp threshold for a universal R\'enyi area law within this adiabatically connected class.
\end{abstract}

\maketitle

\section{Introduction}
\label{sec:main-introduction}

Although the Hilbert space of a quantum many-body system grows exponentially with system size, entanglement in the ground states of noncritical local systems is expected to be strongly constrained. This expectation is embodied in the \emph{entanglement area law}: the von Neumann entropy of a ground state should scale with the boundary of a subsystem rather than its volume~\cite{RevModPhys.82.277}. Beyond characterizing the structure of quantum phases, area laws underpin tensor-network representations of ground states and the numerical study of quantum many-body systems~\cite{PhysRevLett.90.227902,PhysRevB.73.094423,PhysRevLett.100.030504,SCHOLLWOCK201196,RevModPhys.93.045003}.

In one dimension, the area law has been established for general local quantum spin systems with a uniform spectral gap, and the methods of proof and computation have been progressively refined~\cite{Hastings_2007,PhysRevB.85.195145,arad2013area,brandao2013area}. The theory has also been extended to sufficiently rapidly decaying long-range interactions~\cite{Kuwahara2020arealaw,2jgh-nrj1} and broad classes of interacting bosonic systems~\cite{PhysRevA.108.042422,kim2024Boson}. In two and higher dimensions, by contrast, deriving an area law from the spectral gap \emph{alone} has long been a central open problem~\cite{RevModPhys.82.277}. Strong bounds approaching an area law have been obtained for frustration-free systems with a uniform local gap~\cite{Anshu2022Subvolume,10.1145/3519935.3519962}. Other approaches invoke additional input, such as a gapped adiabatic path to a state with known entanglement structure, low-energy densities of states, or thermodynamic properties, and so on~\cite{PhysRevB.72.045141,PhysRevLett.111.170501,Marien2016,PhysRevLett.113.197204,PhysRevB.76.035114,PhysRevA.80.052104,PhysRevB.92.115134,PhysRevA.66.042327,PhysRevLett.94.060503,PhysRevLett.100.070502,kim2025spectral}. Despite these advances under additional assumptions, it remained unclear how strongly locality and a uniform gap, on their own, constrain entanglement.

Among these approaches, \emph{approximate ground-state projectors} (AGSPs) have provided a particularly systematic link between a spectral gap and entanglement bounds~\cite{PhysRevB.85.195145,arad2013area,Arad2017}. An AGSP $K$ that preserves the ground state exactly contracts the orthogonal excited-state subspace by a factor $\eta<1$ in norm, while having operator Schmidt rank at most $D$ across a specified cut. A standard \emph{good AGSP} condition is, for example, $D\eta^2\le 1/2$. This means that the suppression of excited-state components outweighs the associated growth of Schmidt rank; a bootstrapping argument then yields an entanglement bound of the form $S_1=O(\log D)$. In one dimension, improvements in this competition led to area-law proofs, matrix-product-state approximations, and polynomial-time algorithms~\cite{arad2013area,landau2015polynomial,Arad2017}. Optimizing the \emph{rank--error trade-off}---how small the rank can be made at a prescribed projection accuracy---was therefore a fundamental question in its own right.

The reach of AGSPs extends beyond the von Neumann entropy. For $0<\alpha<1$, the R\'enyi entropy $S_\alpha$ is especially sensitive to small Schmidt weights, and an area law for $S_1$ alone does not imply one for $S_\alpha$. Nevertheless, the standard iteration and bootstrapping of an exactly invariant good AGSP yield $S_\alpha=O(\log D)$ at least for $\alpha\ge 1/2$. More refined control of the approximate Schmidt rank as the error $\epsilon$ tends to zero can constrain the full Schmidt tail, including the regime $0<\alpha<1/2$. In one dimension, such accuracy-dependent estimates imply a R\'enyi area law for every fixed $\alpha>0$~\cite{arad2013area}. This suggests a natural conjecture in higher dimensions: for each fixed $0<\alpha<1$, do all uniformly gapped local lattice Hamiltonians satisfy $S_\alpha(\Omega)\le C_\alpha\mathcal A$? Here $\mathcal A$ is the cross-sectional size of the cut, and $C_\alpha$ may depend on $\alpha$, the fixed local parameters, and the lower bound on the spectral gap, but not on system size. This is the \emph{R\'enyi area-law conjecture} considered in this work.

In two dimensions, however, the competition between AGSP accuracy and rank depends crucially on the boundary size. To illustrate the difficulty, write a representative rank--error trade-off as $\log D_\epsilon\lesssim\mathcal A+(\mathcal A L_\epsilon)^\kappa$, where $L_\epsilon:=\log(100/\epsilon)$. To obtain a good AGSP with a rank cost proportional to the boundary, one must first reach accuracies with $L_\epsilon=\Theta(\mathcal A)$. At that scale, the second term remains $O(\mathcal A)$ only if the exponent satisfies $\kappa\le 1/2$; logarithmic factors must also be controlled at the threshold. This square-root regime is achievable, up to subpolynomial overheads, under stronger assumptions: for locally gapped frustration-free systems~\cite{10.1145/3519935.3519962}.
In one dimension, by contrast, $\mathcal A=O(1)$, and a sublinear accuracy exponent $\kappa<1$ already permits strong control of the Schmidt tail. Earlier one-dimensional AGSP constructions exhibited an accuracy exponent of $\kappa=3/4$~\cite{arad2013area}. Is the failure to obtain the stronger trade-off sought in higher dimensions merely a limitation of current filtering and low-rank techniques? Or is there a \emph{fundamental constraint on the AGSPs permitted by locality and a spectral gap alone}?

Here we answer this question by establishing upper and lower bounds on the AGSP rank--error trade-off. For finite-range local Hamiltonians in any fixed spatial dimension, without assuming frustration freeness or a local spectral gap, we construct AGSPs that approximately preserve the ground state and achieve $\log D_K=\widetilde O\!\left[\mathcal A+(\mathcal A L_\epsilon)^{2/3}\right]$ (Result~1, at a fixed positive physical gap). Conversely, in the high-accuracy regime $L_\epsilon\gtrsim\mathcal A$, we construct explicit Hamiltonian families with a uniform gap and bounded local dimension, interaction range, and interaction strength, for which approximating the ground state or constructing an approximate AGSP requires $\log D=\Omega\!\left[(\mathcal A L_\epsilon)^{2/3}\right]$ (Result~2). In the lower bound, system size is chosen according to the target accuracy. Thus the universal high-accuracy rank--error trade-off is optimal, up to logarithmic factors, at fixed gap. In one dimension, this improves the earlier $3/4$ accuracy exponent to $2/3$ and simultaneously establishes its optimality. 

Crucially, this limitation is not merely a shortcoming of AGSP constructions. The ground states in the same lower-bound construction obey a \emph{volume law}, $S_\alpha(\Omega)=\Omega(n)$: in two dimensions, for each fixed $0<\alpha<1$ one can choose an appropriate family of rectangular lattices, while in three and higher dimensions a single hypercubic family works for every fixed $0<\alpha<1$ (Result~3). Yet these states are connected to product states by paths of local Hamiltonians with a uniform spectral gap, and they obey a von Neumann area law $S_1(\Omega)=O(\mathcal A)$ across the same cuts. Our results therefore not only delimit what can be achieved by universally optimizing AGSPs, but also disprove the universal R\'enyi area-law conjecture itself for every $0<\alpha<1$. Unlike previously known examples of large entanglement in one-dimensional chains whose gap closes~\cite{GottesmanHastings2010,Irani2010,Bravyi2012Motzkin,MovassaghShor2016,Zhang2017Motzkin}, or counterexamples to generalized area laws in less restrictive interaction geometries~\cite{aharonov2014local}, our construction retains both a local lattice geometry and a physical gap independent of system size.

The mechanism behind this separation is what we call \emph{error-space complexity}: large entanglement hidden in components of exceptionally small weight. That small Schmidt weights can separate R\'enyi entropy scaling from MPS approximability is already known~\cite{PhysRevLett.100.030504}, and related uses of local flags and dynamical embeddings have precedents~\cite{PhysRevLett.110.060504,Mackay2026OptimalCompression}. The essential feature here is that a highly entangled component can be extracted by an \emph{a priori physically specified product projector across the cut}, so its Schmidt tail yields a lower bound on the tail of the full state~\cite{PhysRevLett.83.1046}. We implement this idea using parallel Feynman--Kitaev-type history states~\cite{Feynman1985,KitaevShenVyalyi2002}, maintaining a uniform physical gap through a single root well shared by the rows while retaining a small-weight component containing many Bell pairs. This explains the common origin of the matching AGSP lower bound and the R\'enyi volume law.

\section{Setup}
\label{sec:main-setup}

We consider a quantum spin system on a finite rectangular lattice $\Lambda$ with open boundary conditions, in a fixed spatial dimension $\nu\ge1$ and with on-site Hilbert-space dimension at most $d_0$. The upper bound also applies to the more general layered geometries specified in Appendix~\ref{app:upper-finish}.

Throughout, $\|\cdot\|$ denotes the Hilbert-space norm for vectors and the operator norm for operators; $\|X\|_1=\operatorname{tr}\sqrt{X^\dagger X}$ denotes the trace norm.
The Hamiltonian is finite-range and $k$-local,
\begin{align}
H=\sum_{\substack{Z\subseteq\Lambda:\,1\le |Z|\le k\\ \diam(Z)\le r_0}}h_Z,
\qquad
\max_{i\in\Lambda}\sum_{Z\ni i}\|h_Z\|\le1,
\label{eq:main-setup-Hamiltonian}
\end{align}
where $h_Z=h_Z^\dagger$ and $d_0,k,r_0$ are independent of the system size. We assume a unique normalized ground state $\ket{\Omega}$ with ground energy $E_0$ and spectral gap $\Delta>0$. No frustration-free or local-gap assumption is imposed.

Fix a bipartition $\Lambda=L\sqcup R$ by a plane between two adjacent layers, and let $\mathcal{A}$ denote the number of sites in one cross-section of the cut. For a rectangular box cut perpendicular to the first coordinate, $\mathcal{A}=\prod_{j=2}^{\nu}N_j$; in one dimension $\mathcal{A}=1$.

For a pure state or an operator, $\SR$ denotes the Schmidt rank or operator Schmidt rank across this fixed cut. We also use the approximate Schmidt rank
\begin{align}
\begin{gathered}
D_\epsilon(\Omega):=\min\bigl\{\SR(\phi):\ \|\ket{\Omega}-\ket{\phi}\|\le\epsilon\bigr\} 
\end{gathered}
\label{eq:main-setup-approx-rank}
\end{align}
with $0<\epsilon<1$,
where $\ket{\phi}$ need not be normalized. For $\rho_L=\tr_R\ket{\Omega}\bra{\Omega}$, the R\'enyi entanglement entropy is
$S_\alpha(\Omega)=(1-\alpha)^{-1}\log\tr(\rho_L^\alpha)$ for $0<\alpha<1$; its continuous extension at $\alpha=1$ is the von Neumann entropy.

An approximate ground-state projector (AGSP) is a Hermitian operator $K$ for which there is a normalized state $\ket{\Omega_K}$ satisfying
\begin{align}
&K\ket{\Omega_K}=\ket{\Omega_K},\qquad \SR(K)\le D_K, \notag \\
&\|\ket{\Omega}-\ket{\Omega_K}\|\le\delta_K, \notag \\
&\bigl\|K(1-\ket{\Omega_K}\bra{\Omega_K})\bigr\|\le\epsilon_K.
\label{eq:main-setup-AGSP}
\end{align}
We write $\eta_K:=\delta_K+\epsilon_K$. Then $\|K-\ket{\Omega}\bra{\Omega}\|\le\eta_K$. The exactly invariant case corresponds to $\delta_K=0$.
Unlike the conventional exactly invariant AGSP, our definition allows a small deviation of the invariant state from the true ground state~\cite{Kuwahara2020arealaw,kim2024Boson}. Our rank-error optimality statement refers to this approximate notion.

\section{Error-space complexity}
\label{sec:error-space-complexity}

\begin{figure*}[!t]
  \centering
  \includegraphics[width=\textwidth]{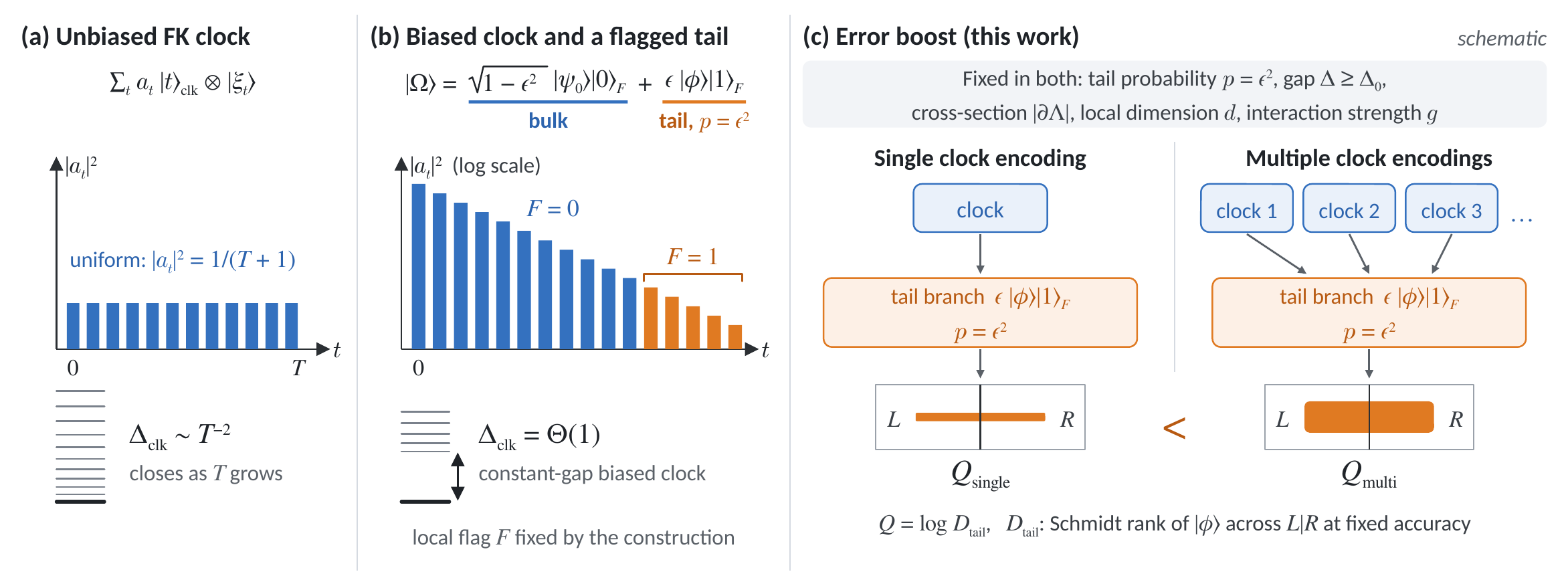}
  \caption{\textbf{Error-space complexity in a flagged component.}
  (a) A uniform Feynman--Kitaev history has equal clock probabilities and
  a clock gap that decreases as $T^{-2}$ for fixed local normalization.
  (b) A biased clock illustrates how a finite gap can coexist with a
  low-weight, physically distinguishable tail. The flag outcome $F=1$
  selects the normalized component $|\phi\rangle$ with probability
  $p=\epsilon^2$, as in Eq.~\eqref{eq:error-space-flag}.
  (c) At the same tail probability and local resource bounds,
  $Q=\log D_{\rm tail}$ is the logarithm of the Schmidt rank needed
  to approximate the normalized tail at fixed accuracy. The two occurrences of
  $|\phi\rangle$ denote the respective tails, not an identical state.
  The comparison is schematic: in our construction, exchange orders merge
  at common configurations and parallel rows share one attractive root,
  rather than being independent clock copies (Appendices~\ref{app:row}
  and~\ref{app:parallel}).
  The clock gaps in (a,b) refer to the reference clock models; the
  fixed $\Delta$ in (c) is the full physical ground-state gap.
  The latter does not require a nonvanishing internal gap of $J_b$.
  Here $\mathcal A$ is the cross-sectional size, $d$ the local dimension,
  and $g$ the local-strength bound; in Eq.~\eqref{eq:main-setup-Hamiltonian},
  $d\le d_0$ and $g$ is normalized to $1$.}
  \label{fig:error-space}
\end{figure*}

Small-weight components with large entanglement have long been used to
separate R\'enyi entropy scaling from MPS approximability~\cite{PhysRevLett.100.030504}.
Local flags have also been used to embed universal quantum computation into
states close to product states~\cite{PhysRevLett.110.060504}.
R\'enyi-index-dependent changes in the entanglement of dominant Schmidt
sectors have also been studied in locally quenched
dynamics~\cite{Grover2026Hierarchical}.
Our lower bounds apply this perspective to uniformly gapped ground states,
adapting its use in quantum dynamics~\cite{Mackay2026OptimalCompression}.
Schematically, for a decomposition
$\ket{\psi}=a\ket{\psi_0}+\epsilon\ket{\phi}$,
we use \emph{error-space complexity} to describe how complex the normalized component
$\ket{\phi}$ can be as a function of its amplitude $\epsilon$.
Here, the relevant measure is the Schmidt rank required to approximate this
component to fixed constant vector-norm accuracy $\delta$ across $L|R$;
$\epsilon$ denotes the branch amplitude, not this approximation tolerance.
Approximate Schmidt rank is a standard quantity in one-shot entanglement
theory~\cite{PhysRevLett.106.130503}; here we keep the physical extraction
map and its success probability explicit. A complementary measure of the
width of the Schmidt spectrum is smooth entanglement spread, defined through
differences of smooth extremal entropies~\cite{Anshu2022EntanglementSpread}.

An essential requirement is that the two components can be separated by a
product projection fixed \emph{a priori} by the physical construction,
independently of the chosen approximation. Here ``locally identifiable'' means local with respect to the bipartition: the extraction is $P_L\otimes P_R$. Its support and circuit depth within either half need not be bounded, and its success probability can be very small.
A simple realization uses a local flag qubit $F$:
\begin{align}
\ket{\psi}
=\sqrt{1-\epsilon^2}\ket{\psi_0}\ket{0}_F
+\epsilon\ket{\phi}\ket{1}_F,
\label{eq:error-space-flag}
\end{align}
where both component states are normalized, $0<\epsilon<1$, and $F$ lies
entirely on one side of the cut.
The flag outcome $1$ has probability $p=\epsilon^2$.
Then $P=I\otimes\ket{1}\bra{1}_F$ is a product projector across the cut and
extracts $\epsilon\ket{\phi}\ket{1}_F$.
More generally, for a fixed product projector $P$, write
$p=\|P\ket{\psi}\|^2$ and
$\ket{\psi_P}=P\ket{\psi}/\sqrt p$.
At fixed $0<\delta<1$, the error-space complexity is
$Q_\delta(\psi;P):=\log D_\delta(\psi_P)$, considered together with its
sector probability $p$ (Fig.~\ref{fig:error-space}c).
The single-qubit flag is illustrative: in our Hamiltonian construction,
$P$ tests activation and particle-number conditions and may have extensive
support within one side of the cut.
Without a fixed product projection, unrestricted decompositions can assign highly
complex error components even to a product state, through cancellation
between the two terms.

This separation makes the tail useful for lower bounds.
Since product projection is contractive and cannot increase Schmidt rank,
the ordered Schmidt coefficients $\{\lambda_j\}$ of $\ket{\psi}$ and
$\{\mu_j\}$ of $\ket{\psi_P}$ satisfy
\begin{align}
\sum_{j>D}\lambda_j^2
\ge p\sum_{j>D}\mu_j^2.
\label{eq:error-space-tail-lower-bound}
\end{align}
Indeed, projecting any rank-$D$ approximation cannot increase its rank or
its vector-norm error, so the inequality follows from optimal Schmidt
truncation; it is also the standard stochastic local-monotonicity inequality
for Schmidt-tail sums~\cite{PhysRevLett.83.1046}.
Equivalently, $D_{\delta\sqrt p}(\psi)\ge D_\delta(\psi_P)$:
the full state inherits the tail's rank cost at error scale $\delta\sqrt p$
($=\delta\epsilon$ in the flagged example).
For the AGSP lower bound, our construction also provides a product root
with constant overlap with $\ket{\Omega}$, allowing this state-rank bound
to transfer to the operator Schmidt rank of $K$.
Figure~\ref{fig:error-space} summarizes the distinction between the
weight of a component and the complexity of its normalized state.

\section{Main results}
\label{sec:main-results}
Our results determine the optimal AGSP rank--error trade-off and show why it does not enforce a R\'enyi area law below $\alpha=1$. Let $\mathcal A$ be the cross-sectional size of a prescribed planar cut, $\Delta$ the ground-state gap, and $\epsilon$ the target accuracy. We write $L_\epsilon=\log(100/\epsilon)$ and $\Delta_*=\min\{\Delta,1\}$.

\mainresult{result:main-upper}{AGSP upper bound}
For any finite-range Hamiltonian in a fixed spatial dimension with a unique ground state and a spectral gap $\Delta>0$, there exists, across any prescribed planar cut, an AGSP with error $\eta_K\le\epsilon$ and operator Schmidt rank satisfying
\begin{equation}
\log D_K=\widetilde O\!\left[\mathcal A+
\frac{(\mathcal A L_\epsilon)^{2/3}}{\Delta_*^{1/3}}\right]
\label{eq:main-upper-bound}
\end{equation}
This result requires neither frustration freeness nor a local-gap assumption and applies to general frustrated Hamiltonians; see Appendix~\ref{app:upper} for the proof.

In one dimension, earlier AKLV bounds~\cite{arad2013area} gave a high-accuracy rank cost of $\widetilde O(\Delta^{-1/3}L_\epsilon^{2/3})$ for frustration-free Hamiltonians. For general frustrated systems, the corresponding cost was $\widetilde O(\Delta^{-1/4}L_\epsilon^{3/4})$. Our result achieves the $2/3$ accuracy exponent without frustration freeness and extends to every fixed spatial dimension. At fixed positive gap, it matches the lower bound below up to logarithmic factors. Optimality of the gap dependence is not asserted.

\mainresult{result:main-lower}{Optimality of the AGSP rank--error trade-off}
We next show that the $2/3$ exponent in Result 1 cannot be improved in the worst case. We construct families of finite-range Hamiltonians with fixed local dimension, bounded interaction strength, unique ground states, and a spectral gap bounded below independently of system size.

For a prescribed target accuracy $0<\epsilon\le1/8$ in the high-accuracy regime $L_\epsilon\gtrsim\mathcal A$, their ground states obey
\begin{equation}
\log D_\epsilon(\Omega)\ge c(\mathcal A L_\epsilon)^{2/3}
\label{eq:main-approx-rank-lower}
\end{equation}
and, on the same systems, every AGSP with $\eta_K\le\epsilon$ satisfies
\begin{equation}
\log D_K\ge c(\mathcal A L_\epsilon)^{2/3}.
\label{eq:main-AGSP-lower}
\end{equation}
These lower bounds hold in one dimension and in every fixed higher dimension. At fixed gap, they match Result 1 up to logarithmic factors, establishing the optimality of both the $2/3$ accuracy exponent and the joint dependence on boundary size and accuracy in the worst-case AGSP rank--error trade-off. The Hamiltonian size is chosen for the prescribed accuracy.

The key construction achieves an effective parallelization of Feynman--Kitaev history-state operations. Whereas a conventional single clock encodes a sequence of operations, we coordinate multiple local clocks to parallelize the generation of entanglement. The central challenge is to retain Hamiltonian locality and a system-size-independent physical gap while doing so. Rather than simply stacking independent clocks, our construction couples the rows through a shared global root. The same structure admits an embedding in higher spatial dimensions with an arbitrary prescribed cross-section. The construction and the gap proof are given in Appendices~\ref{app:row}--\ref{app:consequences}.

This optimality also exposes a limitation of conventional AGSP-based approaches to higher-dimensional area laws. Such arguments rely on a favorable balance between the shrinking error and the Schmidt-rank cost, typically expressed as $D_K\epsilon_K^2\lesssim1$. In the higher-dimensional worst case, sufficiently accurate projection and an $O(\mathcal A)$ logarithmic rank cost cannot be maintained simultaneously: increasing accuracy eventually incurs a larger rank cost. Optimizing the conventional AGSP rank--error condition alone therefore does not establish a general higher-dimensional area law.

The same one-dimensional lower-bound construction has consequences for finite-temperature quantum correlations~\cite{PhysRevX.11.011047}. For its Gibbs states $\rho_{\beta,m}$ at inverse temperature $\beta$, the R\'enyi entanglement of purification satisfies, for sufficiently large $\beta$ and sufficiently small $0<\alpha<1$,
\begin{equation}
\sup_{m\ge3}E_{p,\alpha}(\rho_{\beta,m})\ge
c\min\!\left\{\beta^{2/3},
\left(\frac{1-\alpha}{\alpha}\right)^2\right\}.
\label{eq:main-thermal-joint}
\end{equation}
In particular, choosing $\alpha=a\beta^{-1/3}$ with sufficiently small fixed $a>0$ yields $E_{p,\alpha}=\Theta(\beta^{2/3})$ along a suitably size-selected family. Moreover, when $\beta=\Theta(\log(1/\epsilon))$, approximating the Gibbs state to trace-norm error $\epsilon$ requires an operator Schmidt rank satisfying
\begin{equation*}
\log D=\Omega\!\left([\log(1/\epsilon)]^{2/3}\right).
\end{equation*}
These results rule out a $\beta^{1/2}$-type improvement, suggested by the diffusive character of imaginary-time evolution, that would hold uniformly over R\'enyi indices for the entanglement of purification. In particular, for $\alpha\asymp\beta^{-1/3}$, no universal temperature exponent smaller than $2/3$ is possible. In this sense the $2/3$ exponent appearing in existing thermal area-law analyses is unavoidable. 
This partially answers the central question of optimal inverse-temperature scaling in thermal area laws raised in Ref.~\cite{PhysRevX.11.011047}.
Because the known $E_{p,\alpha}$ upper bounds also have nontrivial $\alpha$ dependence, however, this does not assert a complete matching of upper and lower bounds in that regime. Our lower bound likewise does not imply a corresponding lower bound on mutual information, whose temperature dependence may still admit improvement.

\mainresult{result:main-renyi}{Breakdown of the R\'enyi area law}
The central physical consequence is a counterexample to the R\'enyi area-law conjecture. For every fixed $0<\alpha<1$ and every spatial dimension $\nu\ge2$, there are families of finite-range Hamiltonians with fixed local dimension, a uniform spectral gap, and unique ground states for which
\begin{equation}
S_\alpha(\Omega)\ge c_\alpha n.
\label{eq:main-renyi-volume}
\end{equation}
Thus the ground-state R\'enyi entanglement follows a volume law rather than an area law.

In two dimensions, a suitable rectangular aspect ratio can be selected for each fixed $\alpha<1$. More strongly, in every dimension $\nu\ge3$, a single hypercubic family, independent of $\alpha$, exhibits the volume law for every fixed $0<\alpha<1$.

Remarkably, these ground states can be connected to product states along uniformly gapped local adiabatic paths. One can therefore evolve continuously from an unentangled product ground state to a R\'enyi volume-law ground state without closing the Hamiltonian gap. At the same time, stability of the von Neumann area law under quasi-adiabatic continuation gives
\begin{equation}
S_1(\Omega)=O(\mathcal A),\qquad
S_\alpha(\Omega)=\Omega(n)\quad(0<\alpha<1).
\label{eq:main-entropy-separation}
\end{equation}
Consequently, neither criticality nor a phase transition is required for a R\'enyi volume law. Even the strong requirement of belonging to the same gapped phase as a product state does not enforce a R\'enyi area law. Von Neumann and low-index R\'enyi entropies therefore capture fundamentally different aspects of the structure of gapped quantum ground states.

This behavior also contrasts with entanglement generated by local real-time dynamics over a system-size-independent time. For example, finite-time evolution of a product state under a general one-dimensional local Hamiltonian preserves an area law for every fixed positive R\'enyi index; finite-depth local circuits have the analogous property in arbitrary dimension. Our construction shows that the extremely small Schmidt-weight sectors of a gapped ground state may contain entanglement complexity not captured by such finite-time local evolutions.

A R\'enyi volume law, however, does not preclude accurate low-rank approximations. Writing $S_1$ for the von Neumann entropy, Schmidt truncation followed by the AGSP of Result 1 gives an approximation $\widetilde\Omega_\epsilon$ of error at most $\epsilon$ such that
\begin{equation}
\log\operatorname{SR}(\widetilde\Omega_\epsilon)
\le2S_1+\widetilde O\!\left[
\mathcal A+(\mathcal A L_\epsilon)^{2/3}\right].
\label{eq:discussion-approximation-rank}
\end{equation}
Hence low-rank approximations remain available when $S_1=O(\mathcal A)$. This is a statement for one prescribed cut, not an efficient construction of a global tensor-network approximation.

The volume law does not arise because most of the ground-state weight is highly entangled. Rather, a physically identifiable sector with exceedingly small probability contains many Bell pairs. This \emph{error-space complexity} is the common mechanism behind both the breakdown of the R\'enyi area law and the optimal AGSP rank--error trade-off.

\section{Outlook}
\label{sec:outlook}
Our rank lower bounds apply in the high-accuracy regime $L_\epsilon\gtrsim\mathcal A$; a matching lower bound at all accuracies remains open. The optimal dependence of the AGSP upper bound on the spectral gap is likewise unknown. Determining the full rank--error--gap trade-off would sharpen our understanding not only of ground-state approximability but also of finite-temperature entanglement and the optimality of thermal area laws.

More fundamentally, our results suggest that the worst-case AGSP rank--error trade-off does not, by itself, contain enough information to prove a general higher-dimensional von Neumann area law. Very large entanglement complexity can reside in Schmidt sectors of tiny weight, while contributing little to the von Neumann entropy. Additional control of the distribution or spatial organization of Schmidt weight may therefore be essential.

Our AGSP lower bound exposes a limitation of the conventional framework, which requires uniform contraction over the entire excited-state subspace, but does not rule out a state-dependent good-AGSP condition \(D_K\epsilon_\psi^2<1\) holding for a particular input state $\ket{\psi}$ in a single application. Indeed, an exponentially small initial ground-state overlap may be amplified to a constant in one step, even though the same contraction condition may fail under subsequent iterations. Such one-shot, state-dependent contraction is compatible with our R\'enyi volume-law counterexamples. Whether adaptive schemes that update the AGSP according to the evolving state can exploit this possibility to achieve efficient ground-state approximation remains an interesting open question.

\medskip
\noindent\emph{Note added.---}
%
While this manuscript was being completed, Ref.~\cite{OpenAI2026AreaLaw} appeared, presenting a proposed proof of a von Neumann entanglement area law for generic two-dimensional gapped local Hamiltonians under a global spectral-gap assumption. Assuming this result, monotonicity of R\'enyi entropies would yield an area law for all $\alpha\ge1$. Our construction proves that no corresponding universal statement is possible for any fixed $0<\alpha<1$. Under this assumption, $\alpha=1$ would therefore be the sharp threshold for a universal R\'enyi area law in two-dimensional gapped systems. Our main results, including the von Neumann area law for the constructed families in Appendix~\ref{app:adiabatic}, are independent of Ref.~\cite{OpenAI2026AreaLaw}.

\section*{Acknowledgements}

This work was supported by the RIKEN Hakubi Fellows Program.
T.~K. is supported by the Japan Science and Technology Agency through its Exploratory Research for Advanced Technology program (JST ERATO Grant Number JPMJER2302) and by JSPS KAKENHI Grant Numbers JP23K25796, JP25K24674, JP26K00019, JP26K00020, and JP26H02015.
T.~K. also would like to thank the Isaac Newton Institute for Mathematical Sciences, Cambridge, for support and hospitality during the programme ``Mathematics of many-body entanglement'' where work on this paper was undertaken. This work was supported by EPSRC grant no EP/Z000580/1.

\subsection*{Acknowledgment of AI assistance}
This project has made use of ChatGPT throughout its development, beginning with GPT-5 in August 2025 and subsequently using successive models in the GPT-5 series through GPT-5.6, as well as GPT-6 Astra. The overall research program, central questions, and guiding conceptual framework were developed by the author. ChatGPT was used extensively for proof exploration, mathematical derivations, consistency checks, and the generation and evaluation of alternative technical approaches. For the lower-bound part of the work, both the author and ChatGPT explored possible constructions. 
Among these, the lower-bound construction establishing the optimal scaling in this manuscript originated from a proposal by GPT-6 Astra. The construction and its rigorous analysis were subsequently developed and refined through extensive collaboration between the author and the model.
The author takes full responsibility for the final statements, proofs, and presentation.

\def\bibsection{\section*{References}} 

\bibliography{high_D_Renyi_cited}

\clearpage
\onecolumngrid
\appendix
\numberwithin{figure}{section}
\renewcommand{\thefigure}{\thesection\arabic{figure}}
\renewcommand{\theHfigure}{appendix.\thesection.\arabic{figure}}

\section{Overview and roadmap}
\label{app:overview}
\label{Sec:method}

This appendix gives the precise statements and proofs underlying
Results~\ref{result:main-upper}--\ref{result:main-renyi} of the main text:
the AGSP upper bound, matching rank lower bounds, and R\'enyi volume laws.
All quantifiers and normalizations are made explicit.

\paragraph{Roadmap.}
Appendix~\ref{app:statements} states the precise rank bounds, and
Appendix~\ref{app:upper} establishes the AGSP upper bound.
Appendices~\ref{app:row} and~\ref{app:parallel} give the local lower-bound
construction and target-weight estimate;
Appendix~\ref{app:consequences} proves the higher-dimensional embedding
and the rank and R\'enyi-entropy consequences.
Appendix~\ref{app:adiabatic} establishes the uniformly gapped path and
the von Neumann area law for our models.
Appendix~\ref{app:thermal} proves the one-dimensional finite-temperature
and Gibbs-compression lower bounds.
The upper and lower bounds follow the complementary routes summarized below.

\paragraph{Upper bound.}
For an arbitrary finite-range Hamiltonian with a unique gapped ground state,
we place a window around the prescribed cut and project only the two outer
regions onto bounded-energy subspaces. A microscopic commutator estimate
controls the discarded ground-state weight. A variational argument then
preserves the ground state and the gap to the required accuracy without
introducing a dependence on the total system size. On the resulting tensor
product space, a short-imaginary-time Chebyshev filter is expanded in
variables attached to the candidate cuts. A binary cut-count decomposition
assigns each low-total-degree term to a cut with few interaction insertions.
A common finite product-operator basis, rather than a pointwise rank bound
inside a time integral, controls the rank of each branch. The two leading
costs are
\begin{align}
 \mathcal A\ell
 \quad\hbox{and}\quad
 \frac{L_\epsilon\sqrt{\mathcal A/\Delta_*}}{\sqrt\ell}.
 \label{app:eq:overview-upper}
\end{align}
Optimizing the window width $\ell$ gives the exponent $2/3$. The proof is
for one prescribed cut; it does not assert an efficiently computable AGSP,
a simultaneous low-rank representation at all cuts, or exact preservation
of the original ground state.

\paragraph{Lower bound.}
The construction starts from an adjacent-exchange graph for a single row.
Its stationary weight depends only on the length of an active interval.
A random-deletion flow bounds the Poincar\'e constant by $O(m^3)$, while
both a product source and a target containing $m$ Bell pairs have stationary
mass bounded below by constants. We realize this graph with six local
states and positive semidefinite local rewrite terms. A local grammar makes
a single attractive well exactly rank one in the legal subspace, yielding
a unique ground state and a constant gap on the \emph{full} physical Hilbert
space, including illegal configurations.

For $M$ rows, one local prefix track activates the rows in sequence.
Its startup cost is $O(M)$, and only the last active row can deactivate.
Consequently, the additional killing rate in the fully active sector is at
most $1/2$, independent of $M$. A Poincar\'e--resolvent estimate and a common
imaginary-time integral give a fixed product projection $P_F$ with
\begin{align}
 \|P_F\Omega\|^2
 \ge \exp\!\left[-C\left(M+\sqrt M\,m^{3/2}\right)\right],
 \qquad
 \frac{P_F|\Omega\rangle}{\|P_F\Omega\|}
 \simeq |\Phi\rangle^{\otimes Mm}\otimes|\Xi\rangle.
 \label{app:eq:overview-lower}
\end{align}
Here $\simeq$ denotes equality after isometries acting separately on the two
sides, and $\Xi$ is normalized. Product projection cannot increase Schmidt
rank or singular values. This transfers the Bell factor to Schmidt-tail,
AGSP-rank, and R\'enyi-entropy lower bounds. The extension to higher
dimensions is an explicit cut-preserving serpentine re-embedding, not a new
mixing or gap estimate.

\paragraph{Conventions and organization.}
All logarithms are natural. We use the cross-sectional site count
$\mathcal A$ from the main text, not an interaction-dependent thickened
boundary. Throughout the upper-bound proof and the state-accuracy
corollary,
\begin{align}
 0<\epsilon\le\tfrac14,\qquad
 L_\epsilon:=\log(100/\epsilon),\qquad
 \Delta_*:=\min\{\Delta,1\}.
 \label{app:eq:conventions}
\end{align}
The $\log(1/\epsilon)$ convention in the asymptotic lower bounds differs by
the fixed constant $\log100$; this difference is accounted for explicitly
below. $C,c>0$ may change from line to line. Upper-bound constants depend
only on the fixed local data $(\nu,d_0,k,r_0)$; lower-bound constants are
uniform in $M,m$ and in the serpentine embedding. Auxiliary parameters used
in different proofs are local to their respective sections. In particular,
the lower-bound row length $b=16m$ is unrelated to the upper-bound
boundary strength $b_\partial$.

The accuracy-dependent lower bounds are statements about families whose
size is selected \emph{after} the target accuracy is specified; they are
not divergent lower bounds on the rank of a fixed finite-dimensional
projector.

\clearpage
\section{Precise statements of the results}
\label{app:statements}

\begin{appthm}[AGSP upper bound]
\label{app:thm:upper}
Let $\Lambda$ be a finite rectangular box with open boundary conditions in
fixed dimension $\nu\ge1$, and let
\begin{align}
 H=\sum_{\substack{Z\subseteq\Lambda:\ 1\le|Z|\le k\\
                    \operatorname{diam}(Z)\le r_0}}h_Z,
 \qquad h_Z=h_Z^\dagger,
 \qquad \max_{i\in\Lambda}\sum_{Z\ni i}\|h_Z\|\le1.
 \label{app:eq:local-assumptions}
\end{align}
The on-site dimension is at most $d_0$, and $\nu,d_0,k,r_0$ are fixed.
Assume that $H$ has a unique normalized ground state $\Omega$, energy $E_0$,
and gap $\Delta>0$. Fix a planar cut $L|R$ with cross-section $\mathcal A$.
There are constants $C,C_*$, independent of $|\Lambda|,\mathcal A,
\epsilon,\Delta$, such that for every $0<\epsilon\le1/4$ there exist
$K=K^\dagger$ and a normalized state $\Omega_K$ satisfying
\begin{align}
 K|\Omega_K\rangle&=|\Omega_K\rangle,&
 \|\Omega-\Omega_K\|&\le\delta_K,&
 \|K(1-|\Omega_K\rangle\langle\Omega_K|)\|&\le\epsilon_K,
 \label{app:eq:agsp-definition}\\
 \delta_K+\epsilon_K&\le\epsilon,&
 \operatorname{SR}(K)&\le D_K,&
 \log D_K&\le
 C\left[\mathcal A+
 \left(\frac{\mathcal A L_\epsilon}{\sqrt{\Delta_*}}\right)^{2/3}\right]
 \log^2\!\left(\frac{C_*\mathcal A L_\epsilon}{\Delta_*}\right).
 \label{app:eq:upper-theorem}
\end{align}
In particular, $\|K-|\Omega\rangle\langle\Omega|\|\le\epsilon$.
Neither frustration freeness nor a local-gap assumption is required.
\end{appthm}

\begin{appthm}[Local construction and finite-size Schmidt tail]
\label{app:thm:lower}
There are fixed constants $d_0\le6$, $k\le5$, $r_0\le2$, $\Delta_0>0$,
and $C>0$ with the following properties. For every $m\ge3$, every fixed
$\nu\ge2$, and every rectangular $(\nu-1)$-dimensional cross-section
$\Sigma$ of size $M\ge1$, there is a Hamiltonian satisfying
\eqref{app:eq:local-assumptions} on
\begin{align}
 \Lambda=\{-16m,\ldots,16m\}\times\Sigma,
 \qquad n=M(32m+1),\qquad \mathcal A=M.
 \label{app:eq:lower-geometry}
\end{align}
The cut lies between $x=0$ and $x=1$. The ground state
$\Omega_{M,m}^{(\nu)}$ is unique, the full-system gap is at least
$\Delta_0$, and a product root $o$ obeys
$|\langle o|\Omega_{M,m}^{(\nu)}\rangle|^2\ge1/2$.
There is a fixed projection $P_F=P_L\otimes I_R$, independent of any
approximating state, such that, with
$w_{M,m}:=\|P_F\Omega_{M,m}^{(\nu)}\|^2$,
\begin{align}
 w_{M,m}&\ge
 \exp\!\left[-C\left(M+\sqrt M\,m^{3/2}\right)\right],
 \label{app:eq:target-theorem}\\
 (\mathcal V_L\otimes\mathcal V_R)
 \frac{P_F|\Omega_{M,m}^{(\nu)}\rangle}{\sqrt{w_{M,m}}}
 &=|\Phi\rangle^{\otimes Mm}\otimes|\Xi\rangle,
 \qquad \|\Xi\|=1,
 \label{app:eq:bell-theorem}
\end{align}
where $\mathcal V_L,\mathcal V_R$ are isometries and
$\Phi=(|00\rangle+|11\rangle)/\sqrt2$.
For every integer $1\le D\le2^{Mm}$,
\begin{align}
 \inf_{\operatorname{SR}(\phi)\le D}
 \|\Omega_{M,m}^{(\nu)}-\phi\|^2
 \ge w_{M,m}\left(1-\frac{D}{2^{Mm}}\right).
 \label{app:eq:finite-tail-theorem}
\end{align}
The approximating vector $\phi$ need not be normalized.
In one dimension there is a nearest-neighbor, two-body construction on
$n=32m$ sites, with $\mathcal A=1$, the same constant-gap and root-overlap
properties, and a fixed projection extracting $m$ Bell pairs with
$w_m\ge e^{-Cm^{3/2}}$.
\end{appthm}

\begin{appcor}[Accuracy-dependent rank lower bounds]
\label{app:cor:accuracy}
There are fixed $c,c',C_0,C_1>0$ such that the following hold.
For a prescribed cross-section of size $\mathcal A=M$ and
$L_\epsilon\ge C_0\mathcal A$, choose
\begin{align}
 m=\left\lfloor cL_\epsilon^{2/3}\mathcal A^{-1/3}\right\rfloor\ge3
 \label{app:eq:accuracy-choice}
\end{align}
in Theorem~\ref{app:thm:lower}. The resulting Hamiltonian satisfies
\begin{align}
 \log D_\epsilon(\Omega)\ge c'(\mathcal A L_\epsilon)^{2/3}.
 \label{app:eq:state-accuracy-theorem}
\end{align}
For the AGSP statement, first prescribe $0<\eta\le1/8$ with
$\log(1/\eta)\ge C_1\mathcal A$, and then choose the Hamiltonian using
\eqref{app:eq:accuracy-choice} with $\epsilon'=\sqrt2\eta$.
On this selected system, \emph{every} AGSP with
$\eta_K=\delta_K+\epsilon_K\le\eta$ satisfies
\begin{align}
 \log D_K\ge c'
 \left(\mathcal A\log\frac1\eta\right)^{2/3}.
 \label{app:eq:agsp-accuracy-theorem}
\end{align}
The one-dimensional family gives the same statements with $\mathcal A=1$.
For a common target accuracy $\epsilon$, setting $\eta=\epsilon$ and
choosing the construction for $\epsilon'=\sqrt2\epsilon$ gives both
state and AGSP lower bounds on the same Hamiltonian, after adjusting
the fixed constants; see Appendix~\ref{app:accuracy-proof}.
\end{appcor}

\clearpage
\section{Proof of the upper bound}
\label{app:upper}

The relation between local and global energy distributions was developed in Ref.~\cite{Arad_2016}. The proof below derives the needed microscopic energy tail directly and uses a variational projection argument, rather than importing a spectral comparison for a soft energy cutoff.

\subsection{A layered window and a microscopic energy-tail estimate}
\label{app:upper-window}

Choose $\ell\ge2$ consecutive candidate cuts, including the prescribed
cut, with neighboring cuts separated by a fixed thickness greater than the
interaction range. They give a decomposition
\begin{align}
 \Lambda=L_0\sqcup L_1\sqcup\cdots\sqcup L_\ell,
 \qquad S:=\bigsqcup_{j=1}^{\ell-1}L_j.
 \label{app:eq:window}
\end{align}
Every interior layer contains $O(\mathcal A)$ sites. The exterior regions
$L_0,L_\ell$ can be arbitrarily large. Every local term crosses at most
one candidate cut, and no term meets both exterior regions. The case in
which the window does not fit is treated at the end of the proof.

Write
\begin{align}
 H=H_{L_0}+H_S+H_{L_\ell}+B,
 \qquad
 b_\partial:=\sum_{\substack{Z:\ h_Z\text{ crosses an outer}\\
                                      \text{interface of }S}}\|h_Z\|
 \le C\mathcal A,
 \qquad \|B\|\le b_\partial.
 \label{app:eq:three-regions}
\end{align}
Only the two exterior interfaces enter $B$; hence $b_\partial$ is
independent of $\ell$. For $Y\in\{L_0,S,L_\ell\}$, set
$E_Y=\lambda_{\min}(H_Y)$ and $D_Y=H_Y-E_YI\ge0$. Thus
\begin{align}
 D:=D_{L_0}+D_S+D_{L_\ell}\ge0,
 \qquad (D+B)\Omega=e_0\Omega,
 \qquad e_0:=E_0-E_{L_0}-E_S-E_{L_\ell}.
 \label{app:eq:D-background}
\end{align}
A product of the three regional ground states is a trial state, while
$D\ge0$ gives a lower bound. Therefore $-b_\partial\le e_0\le b_\partial$.
No positive shift of individual interaction terms is needed.

\begin{applemma}[Exponential energy moment]
\label{app:lem:moment}
For $s_0=1/(4k)$,
\begin{align}
 \|e^{s_0D}\Omega\|\le e^{3b_\partial s_0},
 \qquad
 \|\mathbf1_{(T,\infty)}(D)\Omega\|
 \le e^{-s_0T+3b_\partial s_0}\quad(T\ge0).
 \label{app:eq:moment}
\end{align}
\end{applemma}
\begin{appproof}
Up to a scalar, $D$ is a sub-sum of the original microscopic local terms,
so its interaction strength per site is still at most one. Write
$\operatorname{ad}_A(B):=[A,B]=AB-BA$ and
$\operatorname{ad}_A^0(B):=B$. If $O_Z$ is supported on at most $k$ sites, then
\begin{align}
 \|\operatorname{ad}_D^r(O_Z)\|
 \le(2k)^r r!\,\|O_Z\|,\qquad r\ge0.
 \label{app:eq:microscopic-commutator}
\end{align}
Indeed, after $j$ nonzero commutators the support has at most
$k+j(k-1)$ sites. The sum of the norms of terms that can contribute to the
next commutator is at most $k+j(k-1)\le k(j+1)$. Summing over ordered
sequences of local terms and using $\|[X,Y]\|\le2\|X\|\|Y\|$ yields
$\prod_{j=0}^{r-1}2k(j+1)=(2k)^rr!$.
Applying this estimate term by term to $B$ gives, for $0\le s\le s_0$,
\begin{align}
 \|e^{sD}Be^{-sD}\|
 \le\sum_{r\ge0}\frac{s^r}{r!}
                  \|\operatorname{ad}_D^r(B)\|
 \le\frac{b_\partial}{1-2ks}\le2b_\partial.
 \label{app:eq:conjugated-boundary}
\end{align}
Let $F(s)=e^{sD}e^{-s(D+B)}$. Differentiation gives
$F'(s)=-(e^{sD}Be^{-sD})F(s)$ and $F(0)=I$, so the integral equation and
Gronwall's inequality imply $\|F(s)\|\le e^{2b_\partial s}$.
Since $e^{sD}\Omega=e^{se_0}F(s)\Omega$, the first bound in
\eqref{app:eq:moment} follows. The spectral theorem gives the second.
The argument uses microscopic sites rather than treating an entire layer
as one site; in particular, $s_0$ has no dependence on $\mathcal A$,
$\ell$, or $n$.
\end{appproof}

\subsection{Projection, ground-state stability, and the effective Hamiltonian}
\label{app:upper-projection}

\begin{applemma}[Variational projection lemma]
\label{app:lem:variational}
Let $G\ge0$ have a unique normalized zero-energy ground state $\Omega$ and
gap $\Delta>0$. For an orthogonal projection $P$ with $P\Omega\ne0$, set
\begin{align}
 G_P:=PGP\big|_{\operatorname{Ran}P},
 \qquad
 e_P:=\frac{\langle\Omega|PGP|\Omega\rangle}{\|P\Omega\|^2}.
 \label{app:eq:variational-energy}
\end{align}
If $e_P<\Delta$, the ground state $\Omega_P$ of $G_P$ is unique and,
when embedded in the original space with its phase aligned with $\Omega$,
\begin{align}
 0\le\lambda_0(G_P)\le e_P,
 \qquad \Delta_P\ge\Delta-e_P,
 \qquad \|\Omega_P-\Omega\|\le\sqrt{2e_P/\Delta}.
 \label{app:eq:variational-conclusions}
\end{align}
If $\operatorname{Ran}P$ is one-dimensional, the excited-state assertion
is vacuous.
\end{applemma}
\begin{appproof}
Use $P\Omega/\|P\Omega\|$ as a trial state. For a restriction to a
subspace, the min--max principle gives
\begin{align}
 \lambda_1(G_P)=
 \min_{\substack{U\subseteq\operatorname{Ran}P\\\dim U=2}}
 \max_{\substack{u\in U\\\|u\|=1}}\langle u|G|u\rangle
 \ge\lambda_1(G)=\Delta.
 \label{app:eq:minmax}
\end{align}
This proves uniqueness and the gap estimate. Since
$G\ge\Delta(I-|\Omega\rangle\langle\Omega|)$,
$1-|\langle\Omega|\Omega_P\rangle|^2\le e_P/\Delta$.
For phase-aligned unit vectors, the squared distance is at most twice
this quantity.
\end{appproof}

Project only the two exterior regions:
\begin{align}
 P_L:=\mathbf1_{[0,T]}(D_{L_0}),\qquad
 P_R:=\mathbf1_{[0,T]}(D_{L_\ell}),\qquad
 \Pi:=P_L\otimes I_S\otimes P_R,
 \qquad \mathcal H_\Pi:=\operatorname{Ran}\Pi.
 \label{app:eq:outer-projection}
\end{align}
The three nonnegative summands in $D$ commute, and hence
\begin{align}
 Q_\Pi:=I-\Pi\le\mathbf1_{(T,\infty)}(D),
 \qquad q_T:=\|Q_\Pi\Omega\|
 \le e^{-s_0T+3b_\partial s_0}.
 \label{app:eq:projection-tail}
\end{align}
With $G=H-E_0=D+B-e_0\ge0$, the identities $G\Omega=0$ and
$[D,\Pi]=0$ give
\begin{align}
 0\le\langle\Omega|\Pi G\Pi|\Omega\rangle
 &=-\langle\Pi\Omega|GQ_\Pi\Omega\rangle
  =-\langle\Pi\Omega|BQ_\Pi\Omega\rangle
  \le b_\partial\sqrt{1-q_T^2}\,q_T,\notag\\
 e_\Pi&\le\frac{b_\partial q_T}{\sqrt{1-q_T^2}}.
 \label{app:eq:boundary-only-error}
\end{align}
In particular, the error does not acquire a factor $\|H\|$ or a factor
proportional to the window volume.

Choose the explicit threshold
\begin{align}
 T:=3b_\partial+4k\log\!\left(
            \frac{1024(1+b_\partial)}{\epsilon^2\Delta_*}\right).
 \label{app:eq:threshold}
\end{align}
Then
\begin{align}
 q_T\le\frac{\epsilon^2\Delta_*}{1024(1+b_\partial)}<\tfrac12,
 \qquad e_\Pi\le2b_\partial q_T
                   \le\frac{\epsilon^2\Delta_*}{512}.
 \label{app:eq:projection-budget}
\end{align}
Set
\begin{align}
 H_\Pi:=\Pi H\Pi\big|_{\mathcal H_\Pi},\qquad
 E_\Pi:=\lambda_{\min}(H_\Pi),\qquad
 \widetilde H:=H_\Pi-E_\Pi I_{\mathcal H_\Pi}.
 \label{app:eq:effective-H}
\end{align}
Lemma~\ref{app:lem:variational} shows that $\widetilde H$ has a unique
normalized ground state $\widetilde\Omega$ and satisfies
\begin{align}
 \widetilde H\ge0,\quad
 \widetilde H\widetilde\Omega=0,\quad
 \widetilde\Delta\ge\Delta/2,\quad
 \|\Omega-\widetilde\Omega\|\le\epsilon/16.
 \label{app:eq:effective-properties}
\end{align}
The width of the spectrum of $D_S$ is at most
$2\sum_{Z\subseteq S}\|h_Z\|\le C\mathcal A\ell$.
On $\mathcal H_\Pi$, the two exterior energies are at most $T$, whereas
the compressed $B$ has norm at most $b_\partial$. Consequently,
\begin{align}
 \|\widetilde H\|
 \le2T+C\mathcal A\ell+2b_\partial
 \le C\bigl[\mathcal A\ell+L_\epsilon+\log(1/\Delta_*)\bigr].
 \label{app:eq:effective-norm}
\end{align}
The construction is projection-based; no spectral comparison for a
soft energy cap is used.

\subsection{Crossing ranks and zero extension}
\label{app:upper-compression}

The effective space retains the tensor product structure
\begin{align}
 \mathcal H_\Pi=(\operatorname{Ran}P_L)\otimes\mathcal H_{L_1}
 \otimes\cdots\otimes\mathcal H_{L_{\ell-1}}
 \otimes(\operatorname{Ran}P_R).
 \label{app:eq:effective-tensor-product}
\end{align}
Let $V_s^{\mathrm{orig}}$ be the sum of terms crossing candidate cut $s$,
and put $V_s=\Pi V_s^{\mathrm{orig}}\Pi|_{\mathcal H_\Pi}$.
Across that cut, compression acts separately on the left and the right.
There are $O(\mathcal A)$ possible local supports meeting an elementary
interface: finite range and fixed dimension bound the number of supports
per site. Each local term has operator Schmidt rank at most $d_0^{2k}$.
Therefore
\begin{align}
 \|V_s\|\le C\mathcal A,\qquad
 \operatorname{SR}_s(V_s)\le C\mathcal A,\qquad
 v:=\sum_{s=1}^\ell\|V_s\|\le C\mathcal A\ell.
 \label{app:eq:crossing-data}
\end{align}
All other compressed terms remain on one side of cut $s$. No microscopic
locality of the exterior compressed operators is assumed.

The inclusion $\mathcal J:\mathcal H_\Pi\hookrightarrow\mathcal H_\Lambda$
factorizes across every candidate cut. Thus zero extension
$\mathcal E(O):=\mathcal J O\mathcal J^\dagger$ satisfies
\begin{align}
 \|\mathcal E(O)\|=\|O\|,\qquad
 \operatorname{SR}_s(\mathcal E(O))\le\operatorname{SR}_s(O),\qquad
 \mathcal E(I_{\mathcal H_\Pi})=\Pi.
 \label{app:eq:zero-extension}
\end{align}
All filters and exponentials below are first formed on $\mathcal H_\Pi$
and only then zero-extended. Filtering $\mathcal E(\widetilde H)$ on the
full space instead would incorrectly preserve the artificial zero-energy
subspace $\Pi^\perp$.

\subsection{Multivariable expansion and the cut-count decomposition}
\label{app:upper-cutcount}

We adapt the differential cut-count method of Ref.~\cite{Mackay2026OptimalCompression}; the coefficient decomposition and the absolute remainder bound needed here are proved below.

On $\mathcal H_\Pi$ define
\begin{align}
 \widetilde H(\lambda)=H_0+\sum_{s=1}^\ell\lambda_sV_s,
 \qquad \widetilde H(\mathbf1)=\widetilde H,
 \qquad \mathscr D=\sum_{s=1}^\ell\partial_{\lambda_s}.
 \label{app:eq:variables}
\end{align}
Both $\Pi$ and the scalar ground-energy shift are kept independent of
$\lambda$. In particular, $H_0\ge-vI$. For an entire function $\Phi$,
write
\begin{align}
 F(\lambda)=\Phi[\widetilde H(\lambda)]
       =\sum_{q\in\mathbb N_0^\ell}F_q\lambda^q,
 \quad |q|=\sum_sq_s,
 \quad \lambda^q=\prod_s\lambda_s^{q_s},
 \quad [e^{\mathscr D}F]_0=F(\mathbf1).
 \label{app:eq:taylor-notation}
\end{align}
Here $q_s$ counts insertions of $V_s$, $[x^p]$ means the Taylor coefficient
at $x=0$, and $[\cdot]_0$ sets all variables to zero. The derivatives
commute; the operators need not commute.

\begin{applemma}[Cut count for an absolutely convergent series]
\label{app:lem:cutcount}
Suppose $F$ is norm-absolutely convergent in a neighborhood of the closed
unit polydisc. For an integer $Q\ge0$, set
\begin{align}
 \delta_Q(F):=\sum_{|q|>Q}\|F_q\|,\qquad
 p_*:=\lfloor2Q/\ell\rfloor,\qquad
 N_{\mathrm{ph}}:=\ell(2Q+1)^{\lceil\log_2\ell\rceil}.
 \label{app:eq:cutcount-data}
\end{align}
There are indexed families $\mathcal Z_s$ of unit phases and coefficients
$C_{s,z}$ such that
\begin{align}
 F(\mathbf1)
 &=\sum_{s=1}^\ell\sum_{z\in\mathcal Z_s} C_{s,z}
       \sum_{p=0}^{p_*}[x^p]
       F(z_1,\ldots,z_{s-1},xz_s,z_{s+1},\ldots,z_\ell)+R_Q,
 \label{app:eq:cutcount}\\
 |z_t|&=1,\qquad |C_{s,z}|\le1,\qquad
 \sum_s|\mathcal Z_s|\le N_{\mathrm{ph}},\qquad
 \|R_Q\|\le2N_{\mathrm{ph}}\delta_Q(F).
 \label{app:eq:cutcount-error}
\end{align}
\end{applemma}
\begin{appproof}
First consider a polynomial of total degree at most $Q$.
Organize the $\ell$ cuts into a binary tree, splitting each set into
children whose cardinalities differ by at most one. For a monomial
$\lambda^q$, compare the \emph{total} costs of the two children $I,J$:
\begin{align}
 d_{I,J}(q):=\sum_{s\in I}q_s-\sum_{s\in J}q_s.
 \label{app:eq:total-cost}
\end{align}
Choose $I$ if $d_{I,J}\le0$ and $J$ otherwise. This tie convention gives
every monomial a unique leaf. The selected cost is at most half the parent
cost at each step. A leaf depth $h$ lies between
$\lfloor\log_2\ell\rfloor$ and $\lceil\log_2\ell\rceil$, so its selected
cut satisfies $q_s\le Q2^{-h}\le2Q/\ell$, and thus $q_s\le p_*$.

Let $N_F=2Q+1$ and $\omega=e^{2\pi i/N_F}$.
The selection functions $\chi_L(d)=\mathbf1_{d\le0}$ and
$\chi_R(d)=\mathbf1_{d>0}$ have the exact discrete Fourier representations
on $-Q\le d\le Q$
\begin{align}
 \chi_b(d)=\sum_{r=0}^{N_F-1}\widehat\chi_b(r)\omega^{rd},
 \qquad
 \widehat\chi_b(r)=\frac1{N_F}\sum_{a=-Q}^Q\chi_b(a)\omega^{-ra},
 \qquad |\widehat\chi_b(r)|\le1.
 \label{app:eq:selection-fourier}
\end{align}
Every coefficient of $q_s$ in $d_{I,J}$ is $0,1$, or $-1$.
Consequently, $\omega^{r d_{I,J}(q)}$ is implemented by multiplying the
variables by unit phases. Multiply the selection functions along a path,
and at its leaf retain only powers at most $p_*$. The assigned monomial
survives exactly once; all other paths cancel in the Fourier sums.
There are at most $N_F^{\lceil\log_2\ell\rceil}$ phase terms per leaf.
Their coefficients are products of the Fourier coefficients and have
modulus at most one. This proves the identity and the branch count for
polynomials. Repeated phase values are kept as separate indexed terms.

For the full series, split $F=F_{\le Q}+F_{>Q}$, and denote the branch
operation by $\mathscr S$. Absolute convergence allows rearrangement.
One branch's leaf-degree truncation retains any monomial at most once,
so
\begin{align}
 \|F_{>Q}(\mathbf1)\|\le\delta_Q(F),\qquad
 \|\mathscr S F_{>Q}\|\le N_{\mathrm{ph}}\delta_Q(F).
 \label{app:eq:cutcount-tail}
\end{align}
Take $R_Q=F_{>Q}(\mathbf1)-\mathscr S F_{>Q}$.
There is no additional factor $p_*+1$ in this tail estimate.
\end{appproof}

For the operator-valued $F$ in \eqref{app:eq:taylor-notation}, a branch is
\begin{align}
 [x^p]\Phi[A_{s,z}+xB_{s,z}],\qquad
 A_{s,z}:=H_0+\sum_{t\ne s}z_tV_t
       =A_L^{s,z}\otimes I+I\otimes A_R^{s,z},\qquad
 B_{s,z}:=z_sV_s.
 \label{app:eq:branch-background}
\end{align}
These backgrounds need not be Hermitian. Uniformly over all branches,
\begin{align}
 \|A_{s,z}\|\le W:=C_W[\mathcal A\ell+L_\epsilon+\log(1/\Delta_*)],
 \qquad \|B_{s,z}\|\le C\mathcal A\le W,
 \qquad \operatorname{SR}_s(B_{s,z})\le C\mathcal A.
 \label{app:eq:branch-norms}
\end{align}

\subsection{A Gibbs filter with an absolute Taylor-tail bound}
\label{app:upper-filter}

Set $\gamma=\Delta_*/2$ and choose
$a=C_a\mathcal A\ell\ge\max\{8v,1\}$ with $C_a$ sufficiently large.
For the first-kind Chebyshev polynomial $T_{m_f}$, define
\begin{align}
 f_{m_f}(x):=
 \frac{T_{m_f}(2e^{(\gamma-x)/a}-1)}
      {T_{m_f}(2e^{\gamma/a}-1)},
 \label{app:eq:filter}
\end{align}
which relies on a similar idea to~\cite{PhysRevLett.116.097202}. 
We have $f_{m_f}(0)=1$. For $x\ge\gamma$, the numerator's argument belongs
to $[-1,1]$. Since $0<\gamma/a\le1/2$,
\begin{align}
 T_{m_f}(2e^{\gamma/a}-1)
 =\cosh\!\left[m_f\operatorname{arcosh}(2e^{\gamma/a}-1)\right]
 \ge\tfrac12 e^{c m_f\sqrt{\gamma/a}}.
 \label{app:eq:chebyshev-growth}
\end{align}
Taking
\begin{align}
 m_f:=\left\lceil C_mL_\epsilon\sqrt{a/\gamma}\right\rceil
       =O\!\left(L_\epsilon\sqrt{\mathcal A\ell/\Delta_*}\right)
 \label{app:eq:filter-degree}
\end{align}
with fixed $C_m$ sufficiently large gives, on $\mathcal H_\Pi$,
\begin{align}
 \|f_{m_f}(\widetilde H)-\widetilde P_0\|\le\epsilon/16,
 \qquad \widetilde P_0:=|\widetilde\Omega\rangle
                            \langle\widetilde\Omega|.
 \label{app:eq:filter-error}
\end{align}
Notice that the spectral upper edge of $\widetilde H$ does not enter the
filter scale $a$.

As a polynomial in $y=e^{-x/a}$, the filter has the form
\begin{align}
 f_{m_f}(x)=\sum_{j=0}^{m_f}c_j e^{-jx/a},
 \qquad \sum_{j=0}^{m_f}|c_j|\le C_c^{m_f}
 \label{app:eq:filter-coefficients}
\end{align}
for a fixed $C_c>1$. To see this, apply
$T_{r+1}(u)=2uT_r(u)-T_{r-1}(u)$ to the coefficient $\ell^1$ norm after
substituting $u=2e^{\gamma/a}y-1$. The coefficient norm of this linear
polynomial is uniformly bounded, and the denominator in
\eqref{app:eq:filter} is at least one. The recurrence therefore gives a
geometric upper bound in $m_f$.

Take $F(\lambda)=f_{m_f}[\widetilde H(\lambda)]$. For $t=j/a$, let
$E_{j,q}$ be the Taylor coefficients of
$e^{-t\widetilde H(\lambda)}$. Duhamel expansion with $r$ insertions has
simplex volume $t^r/r!$. Since $H_0\ge-vI$, the product of the norms of
the background propagators is at most $e^{tv}$. Summing over all ordered
insertion sequences yields
\begin{align}
 \sum_{|q|=r}\|E_{j,q}\|\le e^{tv}\frac{(tv)^r}{r!},
 \qquad
 \sum_q2^{|q|}\|E_{j,q}\|\le e^{3jv/a}.
 \label{app:eq:absolute-duhamel}
\end{align}
It follows that
\begin{align}
 \sum_q2^{|q|}\|F_q\|\le e^{Cm_f},
 \qquad \delta_Q(F)\le e^{Cm_f}2^{-Q}.
 \label{app:eq:absolute-filter-tail}
\end{align}
This also supplies the absolute convergence required by
Lemma~\ref{app:lem:cutcount}.

Choose $Q=\lceil C_Qm_f\rceil$, with $C_Q$ a sufficiently large fixed
constant. Then
\begin{align}
 \|R_Q\|\le\epsilon/16,
 \qquad p_*\le Cm_f/\ell.
 \label{app:eq:total-degree-choice}
\end{align}
For completeness, this choice is uniform even after accounting for the
number of Fourier branches. Equation~\eqref{app:eq:filter-degree} implies
$\ell\le Cm_f^2$ and $L_\epsilon\le Cm_f$. Hence
\begin{align}
 \log N_{\mathrm{ph}}
 \le C\log^2(m_f+2)+C\log(C_Q+2)\log(m_f+2).
 \label{app:eq:phase-count-log}
\end{align}
The negative exponent $-C_Qm_f\log2$ in
\eqref{app:eq:absolute-filter-tail} dominates $Cm_f$,
$\log N_{\mathrm{ph}}$, and $\log(1/\epsilon)$ for a fixed sufficiently
large $C_Q$. Increasing $C_Q$ increases the branch logarithm only
logarithmically, so there is no circular parameter choice.

\subsection{A common finite product basis for few insertions}
\label{app:upper-finite-rank}

\begin{applemma}[Finite-rank approximation of a Duhamel coefficient]
\label{app:lem:finite-rank}
Suppose $A_0=A_L\otimes I+I\otimes A_R$, $\|A_0\|\le W$, and
$\operatorname{SR}(B)\le R_B$. For $t\ge0$, $p\ge1$, and $0<\eta<1$,
\begin{align}
 D_{p,t}(A_0,B):=[x^p]e^{-t(A_0+xB)}
 \label{app:eq:Dpt}
\end{align}
has an approximation of operator-norm error at most $\eta$ and Schmidt
rank at most $[R_B(N_{\mathrm{ad}}+2)^2]^p$, where
\begin{align}
 N_{\mathrm{ad}}:=\left\lceil C_N\left[
 (p+1)tW+t\|B\|+\log(p+1)+\log(1/\eta)+1\right]\right\rceil.
 \label{app:eq:ad-degree}
\end{align}
Here $C_N$ is universal and sufficiently large. For $p=0$ the coefficient
has rank one exactly. Neither $A_0$ nor $B$ is required to be Hermitian.
\end{applemma}
\begin{appproof}
Let $B_{A_0}(u)=e^{uA_0}Be^{-uA_0}$. The interaction representation gives
\begin{align}
 D_{p,t}=(-1)^p e^{-tA_0}
 \int_{0\le u_p\le\cdots\le u_1\le t}
 B_{A_0}(u_1)\cdots B_{A_0}(u_p)\,du_1\cdots du_p.
 \label{app:eq:interaction-picture}
\end{align}
Replace every conjugated insertion by
$B_{A_0,N}(u)=\sum_{q=0}^N u^q\operatorname{ad}_{A_0}^q(B)/q!$.
For a decomposition $B=\sum_{\beta=1}^{R_B}B_\beta^L\otimes B_\beta^R$,
\begin{align}
 \operatorname{ad}_{A_0}^q(B)=
 \sum_{\beta=1}^{R_B}\sum_{r=0}^q\binom qr
 \operatorname{ad}_{A_L}^r(B_\beta^L)
 \otimes\operatorname{ad}_{A_R}^{q-r}(B_\beta^R).
 \label{app:eq:common-basis}
\end{align}
All $u\in[0,t]$ therefore use the \emph{same} set of at most
$R_B(N+1)(N+2)/2\le R_B(N+2)^2$ product operators. Expand the product of
$p$ insertions in this fixed set and integrate only its scalar
coefficients. The prefactor $e^{-tA_0}=e^{-tA_L}\otimes e^{-tA_R}$ does
not increase the rank, which proves the stated rank bound. A bound on the
rank at each integration time alone would not suffice.

Since $\|\operatorname{ad}_{A_0}^q(B)\|\le(2W)^q\|B\|$,
\begin{align}
 \sup_{0\le u\le t}\|B_{A_0}(u)-B_{A_0,N}(u)\|
 &\le\|B\|\sum_{q>N}\frac{(2tW)^q}{q!}
 \le\|B\|e^{4tW}2^{-N},
 \label{app:eq:ad-tail}\\
 \max\{\|B_{A_0}(u)\|,\|B_{A_0,N}(u)\|\}
 &\le\|B\|e^{2tW}.
 \label{app:eq:ad-factors}
\end{align}
Telescoping the difference of the $p$-fold products and using the simplex
volume gives
\begin{align}
 \|D_{p,t}-\widetilde D_{p,t}\|
 \le p\frac{(t\|B\|)^p}{p!}\,e^{(2p+3)tW}2^{-N}.
 \label{app:eq:insertion-error}
\end{align}
The bounds $(t\|B\|)^p/p!\le e^{t\|B\|}$ and
$2p+3\le3(p+1)$ show that \eqref{app:eq:ad-degree} makes this at most
$\eta$. When $p=0$, use $e^{-tA_0}$ exactly.
\end{appproof}

Each derivative branch in \eqref{app:eq:cutcount} is a sum
$\sum_{j=0}^{m_f}c_jD_{p,j/a}(A_{s,z},B_{s,z})$.
Assign each such Duhamel coefficient the error budget
\begin{align}
 \eta:=\frac{\epsilon}
 {16N_{\mathrm{ph}}(p_*+1)C_c^{m_f}}.
 \label{app:eq:branch-budget}
\end{align}
Summing over branches and retained insertion counts, with
\eqref{app:eq:filter-coefficients}, gives total approximation error at most
$\epsilon/16$. The preceding estimates imply $\log(1/\eta)\le Cm_f$.
One common degree sufficient for all branches is
\begin{align}
 N_{\mathrm{ad}}:=\left\lceil C_U\left[
       (p_*+1)(m_f/a)W+m_f+1\right]\right\rceil.
 \label{app:eq:common-degree}
\end{align}
This is an explicit choice with both a lower and an upper bound:
\begin{align}
 N_{\mathrm{ad}}\le\operatorname{poly}
       (\mathcal A,\ell,L_\epsilon,\Delta_*^{-1}),
 \qquad
 \log(N_{\mathrm{ad}}+2)
 \le C\log(2+\mathcal A+\ell+L_\epsilon+\Delta_*^{-1}).
 \label{app:eq:degree-polynomial}
\end{align}
There is no dependence on the physical system size $n$.

\subsection{Assembly, Hermitian normalization, and optimization}
\label{app:upper-finish}

Let $\widehat K_\Pi$ be the sum of the approximate branches on
$\mathcal H_\Pi$ and put
$\widehat K=\mathcal J\widehat K_\Pi\mathcal J^\dagger$.
To move a decomposition from a candidate cut to the prescribed cut, at
most $O(\mathcal A\ell)$ sites change sides. Moving $q$ sites multiplies
operator Schmidt rank by at most $d_0^{2q}$, by expanding an operator basis
on those sites. Subadditivity of rank, the branch count, and
Lemma~\ref{app:lem:finite-rank} give
\begin{align}
 \operatorname{SR}(\widehat K)
 \le d_0^{C\mathcal A\ell}
       N_{\mathrm{ph}}(p_*+1)(m_f+1)
       [C\mathcal A(N_{\mathrm{ad}}+2)^2]^{p_*}.
 \label{app:eq:assembled-rank}
\end{align}
The formula also covers $p_*=0$. Thus
\begin{align}
 \log\operatorname{SR}(\widehat K)
 \le C\mathcal A\ell+
 C\left(1+\frac{L_\epsilon\sqrt{\mathcal A/\Delta_*}}{\sqrt\ell}\right)
 \log^2(2+\mathcal A+\ell+L_\epsilon+\Delta_*^{-1}).
 \label{app:eq:preoptimized-rank}
\end{align}

Identify $\widetilde P_0$ with its zero extension, and set
$P_0=|\Omega\rangle\langle\Omega|$. For normalized states,
$\|P_0-\widetilde P_0\|\le\|\Omega-\widetilde\Omega\|$.
Adding the four errors in \eqref{app:eq:effective-properties},
\eqref{app:eq:filter-error}, \eqref{app:eq:total-degree-choice}, and
\eqref{app:eq:branch-budget} yields
\begin{align}
 \|\widehat K-P_0\|\le\epsilon/4=:e.
 \label{app:eq:total-upper-error}
\end{align}
Set $K_0=(\widehat K+\widehat K^\dagger)/2$,
$\lambda=\lambda_{\max}(K_0)$, and $K=K_0/\lambda$.
The spectrum of $K_0$ consists of one eigenvalue
$1-e\le\lambda\le1+e$ and all others in $[-e,e]$.
Since $e\le1/16$, the largest eigenvalue is simple. If its normalized
eigenvector is $\Omega_K$, then
\begin{align}
 \epsilon_K\le\frac{e}{1-e},\qquad
 \|(I-P_0)\Omega_K\|\le\frac{e}{1-e},\qquad
 \delta_K\le\frac{\sqrt2e}{1-e}.
 \label{app:eq:normalization-estimates}
\end{align}
The middle bound follows by projecting
$K_0\Omega_K=\lambda\Omega_K$ onto $P_0^\perp$; the last one uses
phase-aligned unit-vector distance. Hence
\begin{align}
 \delta_K+\epsilon_K
 \le\frac{(1+\sqrt2)\epsilon/4}{1-\epsilon/4}\le\epsilon.
 \label{app:eq:agsp-error-finish}
\end{align}
Hermitianization increases rank by at most a factor of two; scalar
normalization leaves it unchanged. Moreover, with
$P_K=|\Omega_K\rangle\langle\Omega_K|$,
$\|K-P_0\|\le\|K-P_K\|+\|P_K-P_0\|\le\epsilon_K+\delta_K$.

Finally choose
\begin{align}
 \ell:=\max\!\left\{2,
      \left\lceil\left(\frac{L_\epsilon^2}{\mathcal A\Delta_*}
                  \right)^{1/3}\right\rceil\right\}.
 \label{app:eq:optimal-window}
\end{align}
Then
\begin{align}
 \mathcal A\ell+
 \frac{L_\epsilon\sqrt{\mathcal A/\Delta_*}}{\sqrt\ell}
 \le C\left[\mathcal A+
       \left(\frac{\mathcal A L_\epsilon}{\sqrt{\Delta_*}}\right)^{2/3}
       \right].
 \label{app:eq:optimized-cost}
\end{align}
Away from the rounded endpoint the two costs balance. If the optimal
integer width is $O(1)$, then $L_\epsilon^2\le C\mathcal A\Delta_*$ and
both costs are $O(\mathcal A)$. The chosen $\ell$ is polynomial in the
parameters, so the logarithm in \eqref{app:eq:preoptimized-rank} is absorbed
into that of \eqref{app:eq:upper-theorem}.

If the window cannot fit because of an exterior boundary, at least one
side of the prescribed cut contains only $O(\ell)$ layers, hence
$O(\mathcal A\ell)$ sites. Take the exact projector $K=P_0$:
\begin{align}
 \operatorname{SR}(P_0)
 \le d_0^{2\min\{|L|,|R|\}}
 \le d_0^{C\mathcal A\ell},\qquad \delta_K=\epsilon_K=0.
 \label{app:eq:short-system}
\end{align}
The same optimized bound follows. This proves
Theorem~\ref{app:thm:upper}, and thus the upper bound in the main text.

\paragraph{Geometric scope.}
The argument also applies beyond rectangular boxes when a layered window
has $O(\mathcal A\ell)$ sites, moving between its cuts crosses
$O(\mathcal A\ell)$ sites, each interface has interaction strength
$O(\mathcal A)$, every local term crosses at most one candidate cut after
fixed-width coarse graining, and each crossing operator has rank
polynomial in $\mathcal A$. It does not require the norm of an exterior
region's internal Hamiltonian to be $O(\mathcal A)$.

\section{Proof of the lower bound: the one-dimensional row}
\label{app:row}

In the construction below, $b$ denotes the half-row length. We eventually
set $b=16m$ with $m\ge3$. Until the final normalization, all Hamiltonians
are unscaled; dividing by one fixed constant changes their gaps but not
their ground states. We write $-\lambda_*$ for the unscaled ground energy
and reserve $\Delta_0$ for the uniform lower bound on the physical gap.

Local rewrite encodings and highly entangled path-state ground states have antecedents in Refs.~\cite{Bravyi2012Motzkin,MovassaghShor2016,Zhang2017Motzkin}. The construction below instead uses an active-interval weight and a local root well to keep the full physical gap uniform. Biased-clock and modified circuit-to-Hamiltonian methods are discussed in Refs.~\cite{PhysRevA.97.062306,Bausch2018analysislimitations}.

\subsection{The exchange graph and active-interval weights}
\label{app:row-graph}

\begin{figure}[!htbp]
  \centering
  \begin{minipage}{0.98\textwidth}
    \noindent\textbf{(a)}\par\smallskip
    \includegraphics[width=\linewidth]{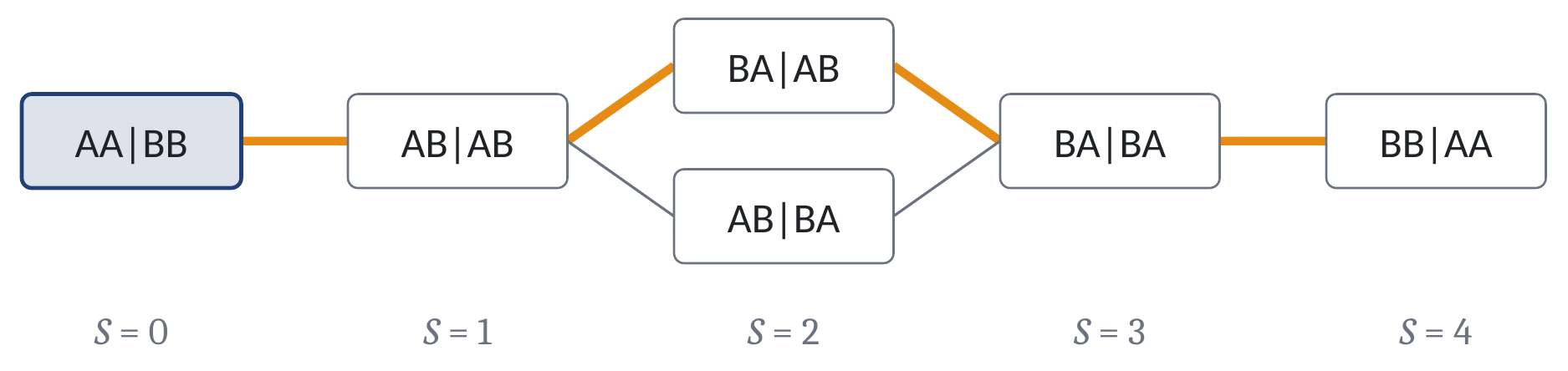}
    \par\medskip\noindent\textbf{(b)}\par\smallskip
    \includegraphics[width=\linewidth]{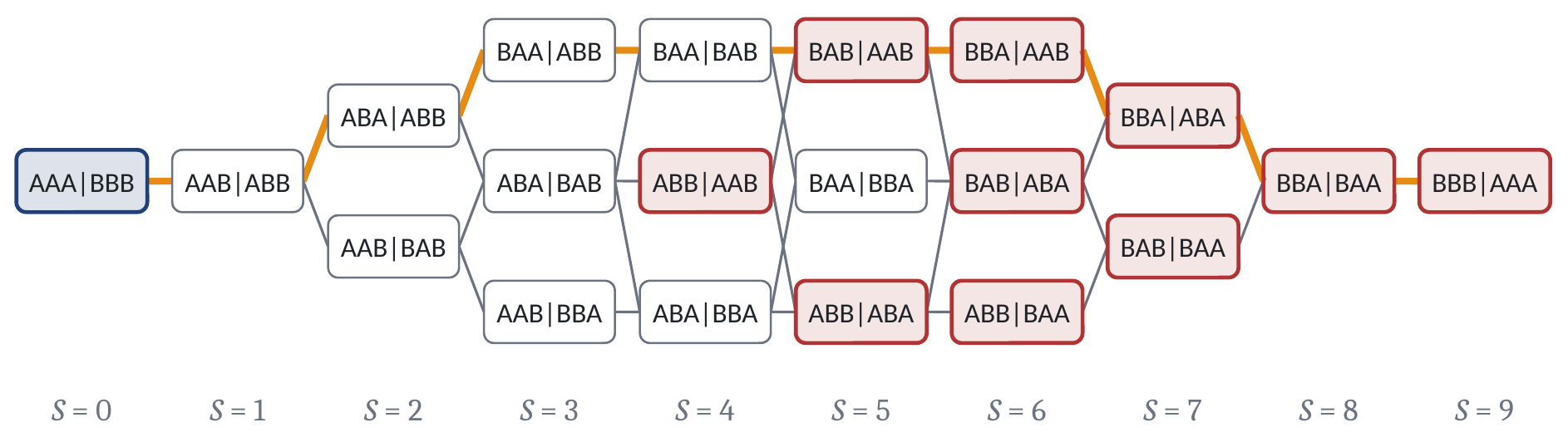}
  \end{minipage}
  \caption{\textbf{Exchange configurations and merged histories.}
  (a) The graph for $a=AA|BB$ has six vertices and six edges. The two
  exchanges available at $AB|AB$ give distinct paths that meet again
  at $BA|BA$; the endpoint is a single vertex, not two separate clock
  states. (b) The graph for $a=AAA|BBB$ has 20 vertices and 30 edges.
  The vertical bar marks the central cut, and columns are indexed by the
  inversion number $S(w)$. Blue marks the root, orange highlights one
  serial exchange path, and red in (b) marks $F_2$, the configurations
  with at least two $B$'s in the left half. All edges represent reversible
  adjacent exchanges $AB\leftrightarrow BA$. The drawing shows adjacency,
  not stationary weights or transition rates; these are defined in
  Eqs.~\eqref{app:eq:active-weight}--\eqref{app:eq:row-rates}.
  These small examples illustrate topology only; the quantitative
  target-mass estimates use $b=16m$, $m\ge3$.}
  \label{fig:app-exchange}
\end{figure}

Let $\mathcal Q_b$ be the set of words containing $b$ copies of $A$ and
$b$ copies of $B$. Edges are adjacent exchanges $AB\leftrightarrow BA$.
The distinguished root is $a=A^bB^b$. Figure~\ref{fig:app-exchange}
shows how different exchange orders merge into a single configuration.
Every nonroot word has a unique
representation
\begin{align}
 w=A^{b-a_u}B\,u\,AB^{b-b_u},\qquad
 a_u=\#A(u)+1,\qquad b_u=\#B(u)+1,
 \label{app:eq:core-word}
\end{align}
where $\#A(u),\#B(u)\le b-1$. The interval $BuA$ extends from the first
$B$ to the last $A$; its core has length $r=|u|\le2b-2$.
The empty core is not the root: it is the root's unique neighbor.
Define
\begin{align}
 \mu(a)=b,\qquad \mu(w)=2^{-|u(w)|}\ (w\ne a),\qquad
 Z_b=\sum_{w\in\mathcal Q_b}\mu(w),\qquad \pi(w)=\mu(w)/Z_b.
 \label{app:eq:active-weight}
\end{align}
For $0\le r\le b-1$, all $2^r$ core words are allowed, so each such
length has unnormalized mass one. The mass at any other length is at most
one. Therefore
\begin{align}
 b\le\sum_{w\ne a}\mu(w)\le2b-1,
 \qquad 2b\le Z_b\le3b-1,
 \qquad \tfrac13\le p:=\pi(a)\le\tfrac12.
 \label{app:eq:root-mass}
\end{align}
For every undirected edge $\{x,y\}$ let
\begin{align}
 h_{xy}:=
 \frac{(\sqrt{\mu(y)}|x\rangle-\sqrt{\mu(x)}|y\rangle)
       (\sqrt{\mu(y)}\langle x|-\sqrt{\mu(x)}\langle y|)}
      {\mu(x)+\mu(y)},
 \qquad J_b:=\sum_{\{x,y\}}h_{xy}.
 \label{app:eq:row-Laplacian}
\end{align}
Each $h_{xy}$ is a norm-one positive projector, and
$J_b\sqrt\pi=0$. The graph is connected because repeated exchanges
$BA\to AB$ reach $a$; hence this zero eigenvector is unique.
The Doob-transformed continuous-time chain has rates and conductances
\begin{align}
 q_{xy}=\frac{\mu(y)}{\mu(x)+\mu(y)},\qquad
 c_{xy}=\pi(x)q_{xy}=c_{yx}.
 \label{app:eq:row-rates}
\end{align}
Exchanges inside the active interval have weight ratio one, extending an
endpoint by one site has ratio $1/2$, and the first root edge has ratio
$1/b$. These are precisely the three types later realized by local rules.

\subsection{A target with stationary mass bounded below}
\label{app:row-target}

For the central cut after the first $b$ letters define
\begin{align}
 F_m:=\{w:\#(B\text{ in the left half})\ge m\}.
 \label{app:eq:row-target}
\end{align}
If $N_A(u)=\#A(u)$, the number of $B$'s on the left is
\begin{align}
 K(w)=1+\#B(u_1\cdots u_{N_A(u)}).
 \label{app:eq:particle-number}
\end{align}
Indeed, the prefix preceding the first $B$ has length $b-N_A(u)-1$,
leaving the first $B$ and the first $N_A(u)$ letters of the core on the
left. Now take $b=16m$. For each $12m\le r\le14m$, we have $r\le b-1$,
so the core conditioned on its length is an independent fair binary word.
The event
\begin{align}
 N_A(u)\ge4m,\qquad \#B(u_1\cdots u_{4m})\ge m
 \label{app:eq:target-event}
\end{align}
implies $K(w)\ge m$.

For completeness, if $X$ is the sum of $r$ independent fair bits, then
$\mathbb E e^{t(X-r/2)}=(\cosh(t/2))^r\le e^{rt^2/8}$.
Exponential Markov inequality, optimized over $t$, gives
$\Pr\{X-r/2\le-a\}\le e^{-2a^2/r}$.
Applying this to the two failures in \eqref{app:eq:target-event}, and using
a union bound, their total probability is at most
$e^{-2m/3}+e^{-m/2}<1/2$ for $m\ge3$.
There are at least $2m$ admissible lengths in this interval, each with
stationary mass $1/Z_b$. Thus
\begin{align}
 q_F:=\pi(F_m)\ge\frac{2m}{2Z_b}\ge\frac1{48}.
 \label{app:eq:target-mass}
\end{align}
The source bound \eqref{app:eq:root-mass} and target bound
\eqref{app:eq:target-mass} concern the same normalized distribution.

\FloatBarrier
\subsection{Random-deletion paths and the Poincar\'e bound}
\label{app:row-poincare}

We use a fractional-path congestion argument in the spirit of multicommodity-flow bounds~\cite{Sinclair1992}; the weight and edge estimates needed here are proved explicitly below.

Orient exchanges toward decreasing inversion number
\begin{align}
 S(w):=\#\{(B,A): B\text{ occurs to the left of }A\}.
 \label{app:eq:inversions}
\end{align}
We have $S(a)=0$ and $S(w)\le b^2$. Given a nonempty core $u$, select
one position uniformly. If the selected letter is $A$, move it to the left
through the core and exchange it with the leading outer $B$. If it is
$B$, move it to the right and exchange it with the trailing outer $A$.
Exchanges of identical letters are omitted. Every actual exchange is
$BA\to AB$, and the resulting core is exactly $u$ with the selected
position deleted. After reaching the empty core, perform the last
$BA\to AB$ exchange to the root. This defines a random path $\gamma_w$
with
\begin{align}
 |\gamma_w|=S(w)\le b^2.
 \label{app:eq:path-length}
\end{align}
All intermediate words remain in $\mathcal Q_b$.

Send mass $\pi(w)$ from every $w\ne a$ along this random path.
Let $F_r(u)$ be the mass reaching core $u$ at the beginning of its
length-$r$ deletion stage. Repeated uniform deletion leaves a uniformly
chosen $r$-subsequence of any starting core of length $R$.
An independent uniform subsequence of a fair binary word is again fair,
and hence
\begin{align}
 \sum_{v\in\{A,B\}^R}2^{-R}
 \Pr\{\text{a uniform }r\text{-subsequence of }v\text{ equals }u\}
 =2^{-r}.
 \label{app:eq:subsequence-identity}
\end{align}
Restricting the sum to allowed starting cores cannot increase it.
Summing over $r\le R\le2b-2$ therefore gives
\begin{align}
 F_r(u)\le(2b-1-r)\frac{2^{-r}}{Z_b}
       \le2b\,\pi(w(u)).
 \label{app:eq:stage-flow}
\end{align}

For a directed edge $e=(z,z')$ toward smaller $S$, define its total
expected traffic by
\begin{align}
 G_e:=\sum_{w\ne a}\pi(w)\,
        \mathbb E[\#\{\text{traversals of }e\text{ in }\gamma_w\}].
 \label{app:eq:edge-traffic-definition}
\end{align}
We claim that
\begin{align}
 G_e\le4b\pi(z),\qquad G_e/c_e\le8b.
 \label{app:eq:congestion-bound}
\end{align}
For an interior exchange in a core of length $r$, consider first a
selected $A$ moving left. Specify its original position $i$ at the start
of the deletion stage. The starting word can then be reconstructed
uniquely: in the core immediately before the exchange at positions
$j,j+1$, remove the $A$ at $j+1$ and reinsert it at $i\ge j+1$.
Omitting swaps of identical letters does not alter this reconstruction
as a word. There are at most $r$ possible values of $i$, each chosen with
probability $1/r$. All reconstructed cores have the same length and
weight as $z$, so \eqref{app:eq:stage-flow} bounds this traffic by
$2b\pi(z)$. For a selected $B$ moving right, remove the $B$ at $j$ and
reinsert it at its original position $i\le j$. The same bound applies.
Together the two possibilities give $4b\pi(z)$.

For an endpoint-shrinking exchange, the same reconstruction removes the
leading $A$ or trailing $B$ of the pre-exchange core and reinserts it at
the selected original position. The same $r$ versus $1/r$ cancellation
applies. The last edge to the root carries the entire nonroot mass;
\eqref{app:eq:stage-flow} at $r=0$ bounds it by
$2b\pi(w(\varnothing))$. This proves the first claim.
In the direction toward the root, the transition rate is $1/2$ for an
interior exchange, $2/3$ for endpoint shrinkage, and $b/(b+1)$ for the
last root edge. Thus $c_e\ge\pi(z)/2$, proving the second claim.

For complex-valued functions $f,g$ on $\mathcal Q_b$, use
$\langle f,g\rangle_\pi=\sum_x\pi(x)\overline{f(x)}g(x)$ and
\begin{align}
 \mathcal E(f,g):=\sum_{\{x,y\}}c_{xy}
       \overline{(f(x)-f(y))}(g(x)-g(y)),\qquad
 \operatorname{Var}_\pi(f):=\|f-\mathbb E_\pi f\|_{L^2(\pi)}^2.
 \label{app:eq:Dirichlet-form}
\end{align}
Each undirected edge is counted once.

\begin{applemma}[Row Poincar\'e estimate]
\label{app:lem:poincare}
The best Poincar\'e constant of the row chain satisfies
\begin{align}
 \operatorname{Var}_\pi(f)\le8b^3\mathcal E(f,f),
 \qquad \tau:=\sup_{\operatorname{Var}_\pi(f)>0}
             \frac{\operatorname{Var}_\pi(f)}{\mathcal E(f,f)}
             \le8b^3.
 \label{app:eq:poincare-bound}
\end{align}
Equivalently, $\tau=1/\lambda_1(J_b)$, where $\lambda_1(J_b)$ is the
smallest positive eigenvalue.
\end{applemma}
\begin{appproof}
Telescoping along each random path and applying Cauchy--Schwarz yields
\begin{align}
 \operatorname{Var}_\pi(f)
 &\le\sum_w\pi(w)|f(w)-f(a)|^2\notag\\
 &\le\sum_w\pi(w)\,
       \mathbb E\!\left[|\gamma_w|\sum_{e\in\gamma_w}|\nabla_e f|^2\right]
 \le b^2\sum_eG_e|\nabla_e f|^2
 \le8b^3\mathcal E(f,f).
 \label{app:eq:path-poincare-proof}
\end{align}
The spectral statement follows from the unitary identification between
$L^2(\pi)$ and the row Hilbert space by multiplication by $\sqrt\pi$.
\end{appproof}

The order $m^3$ can also be checked from below. Every $w\in F_m$ has at
least $m^2$ inversions across the cut, so
$\operatorname{Var}_\pi(S)\ge p q_Fm^4$. An exchange changes $S$ by one,
and the total exit rate at any word is at most $2b-1$. Therefore
\begin{align}
 \mathcal E(S,S)=\sum_{\{x,y\}}c_{xy}
 =\tfrac12\sum_x\pi(x)\sum_yq_{xy}\le(2b-1)/2\le b.
 \label{app:eq:poincare-lower-check}
\end{align}
With $b=16m$, this gives $\tau=\Theta(m^3)$. Here $\tau$ is a relaxation
constant, not a worst-initial-state total-variation mixing time.

\subsection{A six-state local realization}
\label{app:row-local}

Use the on-site alphabet
\begin{align}
 \{U,V,P_0,P_1,T_0,T_1\}.
 \label{app:eq:alphabet}
\end{align}
The symbols $U$ and $V$ denote unvisited exterior sites and vacancies
inside the visited interval. $P_c$ is an ordinary particle with color
$c\in\{0,1\}$, and $T_c$ is a tagged outermost particle. The colors carry
Bell qubits; labels such as $A_i,B_i$ are not stored on physical sites.

\paragraph{Local grammar.}
A legal left half-row is either $U^b$ or
\begin{align}
 U^sT_c v,\qquad 0\le s\le b-1,\qquad
 v\in\{V,P_0,P_1\}^{b-s-1}.
 \label{app:eq:left-grammar}
\end{align}
The right-half grammar is the reflected version. The two central sites
must either both be $U$ or both be non-$U$.
On the left, the permitted nearest-neighbor transitions are
\begin{align}
 U\longrightarrow\{U,T_0,T_1\},\qquad
 T_c\longrightarrow\{V,P_0,P_1\},\qquad
 \{V,P_0,P_1\}\longrightarrow\{V,P_0,P_1\},
 \label{app:eq:grammar-transitions}
\end{align}
and the left endpoint must be $U$ or $T_c$.
The right-hand constraints are their reflection. Let $\Gamma$ be the sum
of the diagonal forbidden-pattern projectors, including the endpoint and
central constraints. Then
\begin{align}
 \ker\Gamma=\mathcal H_{\mathrm{leg}},\qquad
 \Gamma\big|_{\mathcal H_{\mathrm{leg}}^\perp}\ge I.
 \label{app:eq:grammar-gap}
\end{align}
The local central projector identifies the product root exactly inside
the legal space:
\begin{align}
 \Pi_{UU}\big|_{\mathcal H_{\mathrm{leg}}}=|a\rangle\langle a|,
 \qquad |a\rangle=|U\rangle^{\otimes2b}.
 \label{app:eq:physical-root}
\end{align}
Indeed, a central $U$ forces its entire half-row to consist of $U$'s.

\paragraph{Positive local rewrite terms.}
For normalized orthogonal states $\xi,\zeta$ and $\rho>0$, set
\begin{align}
 \Pi_\rho(\xi,\zeta):=
 \frac{(\rho|\xi\rangle-|\zeta\rangle)
       (\rho\langle\xi|-\langle\zeta|)}{1+\rho^2}.
 \label{app:eq:rewrite-projector}
\end{align}
This norm-one projector annihilates $|\xi\rangle+\rho|\zeta\rangle$.
On each internal bond of either half-row, include the color-preserving
hops, and include the indicated tag motion on each respective side:
\begin{align}
 &\Pi_1(VP_c,P_cV) &&\text{on both half-rows},\notag\\
 &\Pi_{1/\sqrt2}(UT_c,T_cV) &&\text{on the left},\notag\\
 &\Pi_{1/\sqrt2}(T_cU,VT_c) &&\text{on the right}.
 \label{app:eq:local-hops}
\end{align}
For each type, the sum over the two colors is a sum of orthogonal blocks
and still has norm one. At the central bond define
\begin{align}
 |\Phi_P\rangle=\frac{|P_0P_0\rangle+|P_1P_1\rangle}{\sqrt2},
 \qquad
 |\Phi_T\rangle=\frac{|T_0T_0\rangle+|T_1T_1\rangle}{\sqrt2},
 \label{app:eq:colored-Bell}
\end{align}
and add
\begin{align}
 h_{\mathrm{pair}}=\Pi_1(VV,\Phi_P),\qquad
 h_{\mathrm{start}}=\Pi_{1/\sqrt b}(UU,\Phi_T).
 \label{app:eq:central-rules}
\end{align}
These act on orthogonal central blocks and each has norm one. Only the
first Bell pair is tagged. Let $J_{\mathrm{phys},b}\ge0$ be the sum of
all rewrite terms.

Every term preserves $\mathcal H_{\mathrm{leg}}$. A hop acts only on
particles and vacancies inward of the tag. Tag motion changes only the
boundary between the exterior $U$ region and the visited interval.
The pair term preserves the requirement that both central sites are
non-$U$. The start term connects the all-$U$ root only to a central tag
pair. Since the terms are Hermitian, $\mathcal H_{\mathrm{leg}}^\perp$
is invariant as well. The coherent pair projectors need not define
unitary-labeled edges on the entire physical space; the exact graph
representation is required only on the invariant history subspace
constructed next.

\subsection{The history isometry and a common Bell factor}
\label{app:row-history}

\begin{figure}[!htbp]
  \centering
  \includegraphics[width=0.96\textwidth]{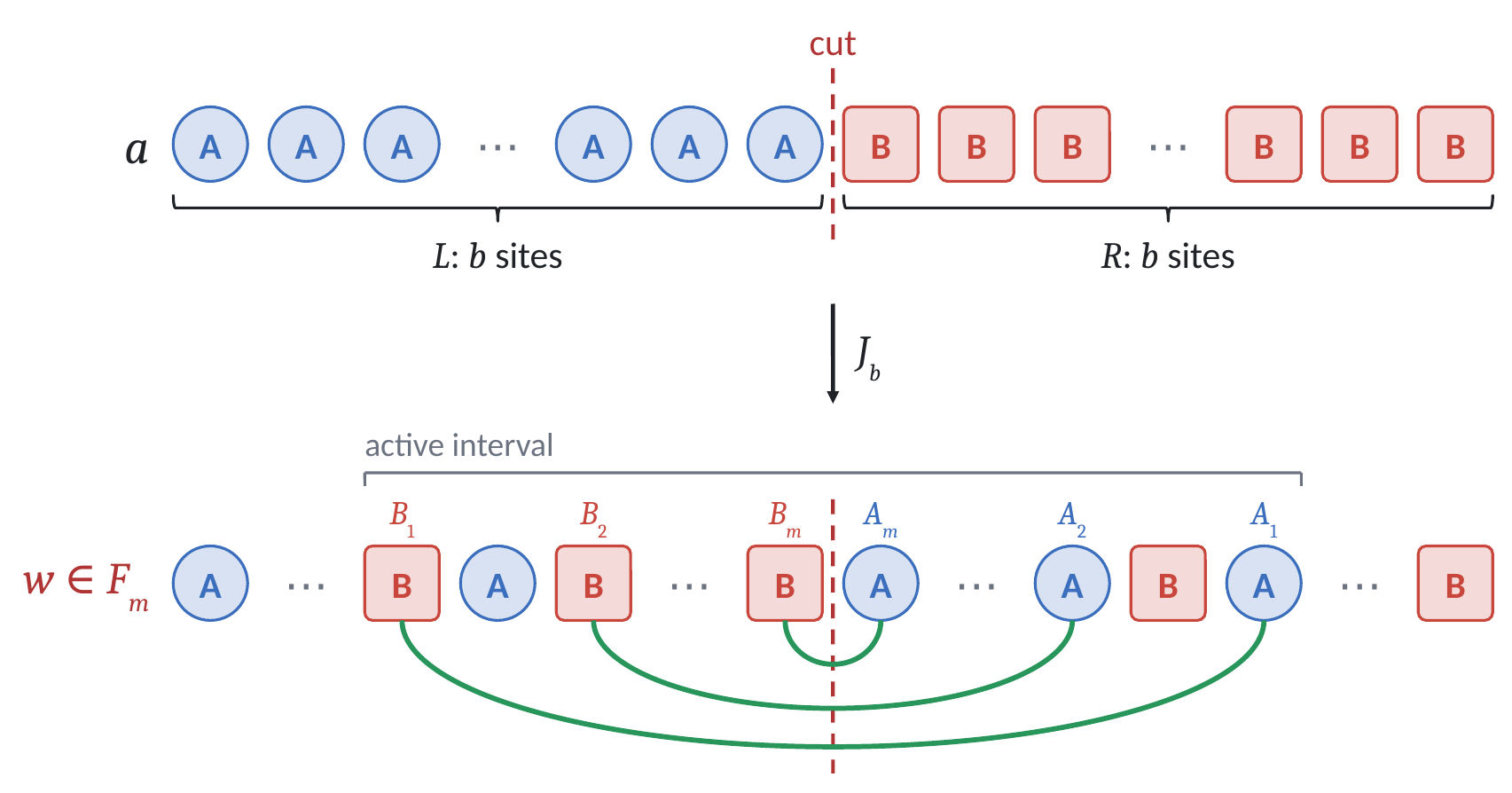}
  \caption{\textbf{One row and its common Bell factor.}
  The upper line is the exchange root $a=A^bB^b$, with $b$ sites on each
  side of the central cut; the lower line is a schematic target
  configuration $w\in F_m$. Ellipses denote omitted sites.
  The $A,B$ symbols label exchange configurations, not the physical
  six-state alphabet. Formal particle labels preserve order within each
  species: only $A_i$ and $B_i$ meet at the central bond, where their
  encoded color registers acquire a Bell pair (green arcs). The outermost
  pair $B_1,A_1$ marks the ends of the active interval. The arrow labelled
  $J_b$ indicates the propagation graph, not a single application taking
  the root to the target. The bound-state amplitudes are proportional to
  $(J_b+\lambda_*)^{-1}|a\rangle$.
  The same outermost $m$ pairs factor from every target superposition:
  $(\mathcal V_L\otimes\mathcal V_R)P_F|\Omega\rangle
   =\sqrt{w_m}\,|\Phi\rangle^{\otimes m}\otimes|\Xi\rangle$.
  The isometries act separately on the two halves, as proved in
  Lemma~\ref{app:lem:history}.}
  \label{fig:app-row}
\end{figure}

Figure~\ref{fig:app-row} illustrates the root, active interval, and Bell pairing.

For a nonroot word $w$, replace its first $B$ and last $A$ by tagged
particles, and all sites outside that interval by $U$. Inside the interval,
replace $A$'s on the left and $B$'s on the right by $V$. The remaining
left $B$'s and right $A$'s become ordinary particles. The two half-rows
have the same particle number $K(w)$. Order the particles on each side
from the outside inward and put a Bell pair between the color registers
with each common rank. Denote the resulting normalized state by
$|\chi_w\rangle$, and let $|\chi_a\rangle=|a\rangle$.
Distinct words have distinct occupation/tag patterns, so the states
$\{\chi_w\}$ are orthonormal.

\begin{applemma}[Exact history representation and Bell extraction]
\label{app:lem:history}
The isometry
\begin{align}
 W:|w\rangle\longmapsto|\chi_w\rangle,
 \qquad
 \mathcal H_{\mathrm{hist}}:=W\mathbb C^{\mathcal Q_b}
 \label{app:eq:history-isometry}
\end{align}
satisfies
\begin{align}
 J_{\mathrm{phys},b}W=WJ_b.
 \label{app:eq:intertwining}
\end{align}
For every normalized vector
$\psi_F\in\operatorname{span}\{\chi_w:w\in F_m\}$ there are fixed
isometries on the two sides such that
\begin{align}
 (\mathcal V_L\otimes\mathcal V_R)|\psi_F\rangle
       =|\Phi\rangle^{\otimes m}\otimes|\Xi\rangle,
 \qquad \|\Xi\|=1.
 \label{app:eq:row-Bell-factor}
\end{align}
The target projection can be chosen as $P_F=P_L\otimes I_R$ by counting
particles in the left half-row.
\end{applemma}
\begin{appproof}
Particles exchange only with vacancies, so their order is preserved.
A central $VV\to\Phi_P$ rewrite adds a pair at the innermost positions
on both sides; its inverse removes that same pair. Both tagged particles
can return to the central bond only after all ordinary pairs have been
removed. Thus each applicable rewrite maps a history vector to the
corresponding adjacent history vector without changing the Bell pairing.
For an interior exchange the amplitude ratio is one, for extension it is
$1/\sqrt2$, and for the first root edge it is $1/\sqrt b$.
These are exactly $\sqrt{\mu(y)/\mu(x)}$ in
\eqref{app:eq:row-Laplacian}, including the diagonal coefficients of each
normalized edge projector. Summing the edge identities proves
\eqref{app:eq:intertwining} and invariance of the history space.

One can also track formal labels by writing the root as
$A_bA_{b-1}\cdots A_1B_1B_2\cdots B_b$ and preserving order within
each species. An adjacent pair $A_i,B_j$ meets at bond $b-i+j$, counted
from the left. It is at the central bond precisely when $i=j$.
This is a mathematical labeling argument, not an enlargement of the
physical alphabet.

For $w\in F_m$, the outermost $m$ particles on each side always carry the
same $m$ Bell pairs. On each half-row, coherently retain the occupation
pattern and move these ordered color registers into fixed output
registers. This operation is an isometry on the target subspace and can
be extended by unused output space. The remaining occupation and color
registers retain orthogonality between different patterns. The same
isometries work for every $w$, so the common $m$-pair factor also separates
from an arbitrary superposition, giving \eqref{app:eq:row-Bell-factor}.
This statement is about vectors in the initialized history space, not
arbitrary physical states with sufficiently many particles. The
isometries are local across $L|R$; no bound on their circuit depth inside
either half is asserted.
\end{appproof}

\FloatBarrier
\subsection{A local attractive well and the gap on the full space}
\label{app:row-gap}

The following observation is useful both for one row and for the parallel
construction.

\begin{applemma}[Rank-one well with an illegal-sector penalty]
\label{app:lem:well}
Let $J_{\mathrm{phys}}\ge0$ preserve a legal subspace and its orthogonal
complement. Let $\Gamma$ vanish on the legal subspace and satisfy
$\Gamma\ge I$ on its complement. Let a projector $R_o$ preserve these
subspaces and restrict to $|o\rangle\langle o|$ on the legal subspace,
where $o$ is a normalized product state. If
$r_o:=\langle o|J_{\mathrm{phys}}|o\rangle\le1/2$, then
\begin{align}
 H_{\mathrm{raw}}=J_{\mathrm{phys}}+2\Gamma-R_o
 \label{app:eq:general-well}
\end{align}
has a unique negative eigenvalue $-\lambda_*$, with
$1-r_o\le\lambda_*\le1$. Its full-system gap is at least $\lambda_*$,
and its ground state obeys
\begin{align}
 |\langle o|\Omega\rangle|^2\ge\lambda_*\ge\tfrac12.
 \label{app:eq:well-overlap}
\end{align}
If an invariant history subspace contains $o$, the ground state belongs
to that history subspace.
\end{applemma}
\begin{appproof}
On the legal subspace, $H_{\mathrm{raw}}$ is a rank-one negative
perturbation of a positive operator. It has at most one negative
eigenvalue: a two-dimensional negative spectral subspace would contain
a nonzero vector orthogonal to $o$, whose expectation of
$J_{\mathrm{phys}}-|o\rangle\langle o|$ is nonnegative, a contradiction.
The root trial state has energy $r_o-1\le-1/2$, so a negative eigenvalue
exists and is simple. Since $H_{\mathrm{raw}}\ge-I$ on the legal space,
$\lambda_*\le1$. On the illegal space,
$J_{\mathrm{phys}}+2\Gamma-R_o\ge I$.
Every other eigenvalue on the full space is therefore nonnegative,
proving the gap assertion. Taking the ground-state expectation gives
$\lambda_*=|\langle o|\Omega\rangle|^2-
\langle\Omega|J_{\mathrm{phys}}|\Omega\rangle
\le|\langle o|\Omega\rangle|^2$.
Finally, on the legal orthogonal complement of an invariant history
space containing $o$, the well vanishes and the Hamiltonian is
nonnegative. Thus the negative eigenstate lies in the history space.
\end{appproof}

For one row take
\begin{align}
 H_b^{(1),\mathrm{raw}}=J_{\mathrm{phys},b}+2\Gamma-\Pi_{UU}.
 \label{app:eq:H-one-row}
\end{align}
The root expectation of $J_{\mathrm{phys},b}$ is $1/(b+1)$, coming only
from the start term. Lemma~\ref{app:lem:well} and
\eqref{app:eq:physical-root} therefore give
\begin{align}
 \frac{b}{b+1}\le\lambda_*\le1,
 \qquad \Delta_{\mathrm{raw}}\ge\lambda_*\ge\tfrac12,
 \qquad \Omega_m^{(1)}\in\mathcal H_{\mathrm{hist}}.
 \label{app:eq:one-row-gap}
\end{align}
This argument includes every color state and every illegal configuration;
it does not use a conserved particle-number sector, since the central
pair terms change particle number. All terms are one-site or
nearest-neighbor two-site terms, with uniformly bounded norms and a
uniformly bounded number of terms at each site. Dividing by a fixed
$g_*>0$ enforces \eqref{app:eq:local-assumptions} while retaining gap at
least $1/(2g_*)$. The small $1/\sqrt b$ amplitude in the start rule is
allowed: the local-strength assumption is an upper bound, not a uniform
positive lower bound on nonzero couplings.

\section{Proof of the lower bound: local parallel activation and target weight}
\label{app:parallel}

Spatially distributed clocks were developed in Refs.~\cite{BreuckmannTerhal2014,GossetTerhalVershynina2015}. Our prefix rule below is tailored to a fixed Bell-pair history and one attractive root. It is not a uniformly gapped simulation of arbitrary circuits; general criticality results for circuit-to-Hamiltonian mappings~\cite{GonzalezCubitt2018} have that broader quantification and additional history-amplitude assumptions.

\subsection{The two-dimensional Hamiltonian and its global root}
\label{app:parallel-local}

\begin{figure}[!htbp]
  \centering
  \begin{minipage}{0.85\textwidth}
    \noindent\textbf{(a)}\par\smallskip
    \includegraphics[width=\linewidth]{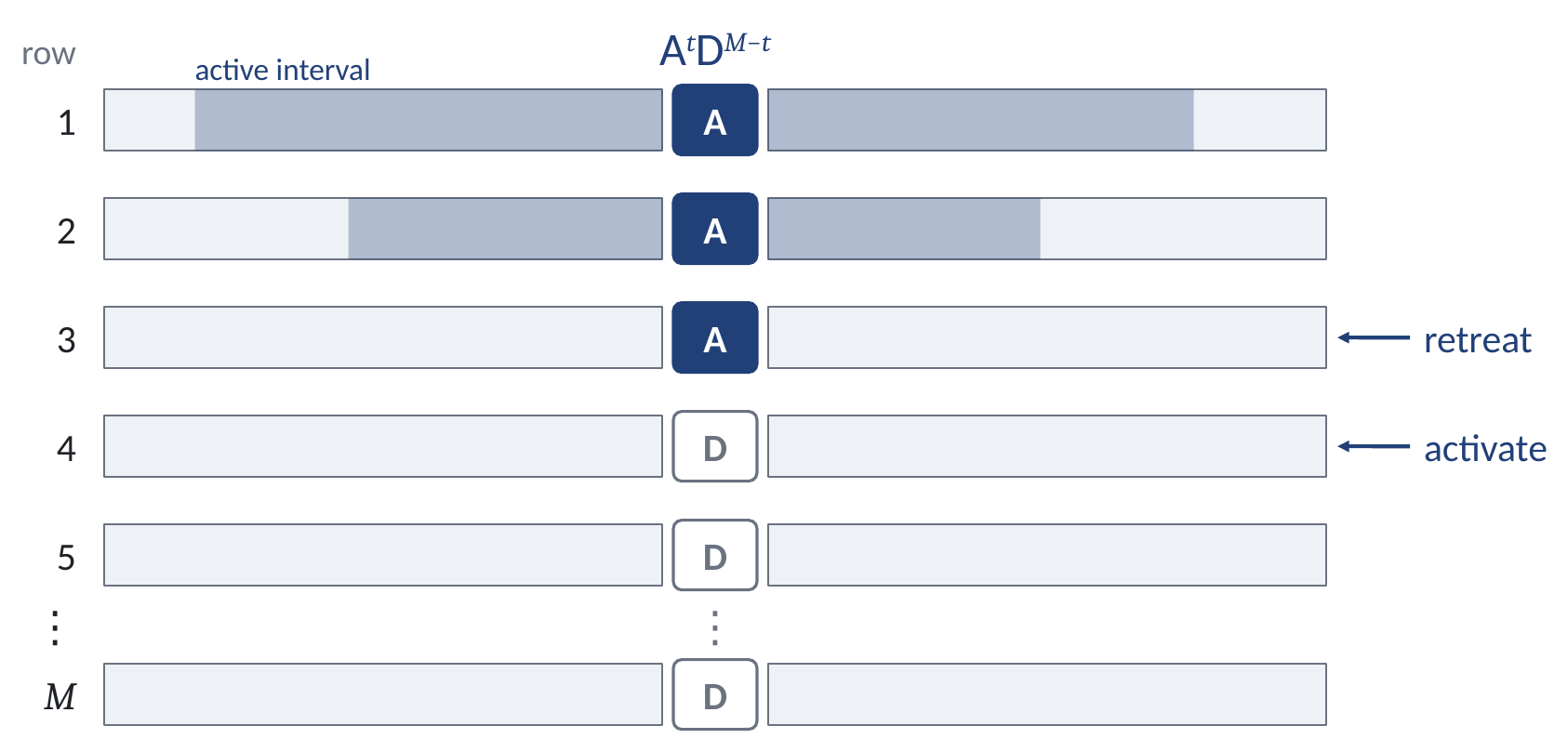}
    \par\medskip\noindent\textbf{(b)}\par\smallskip
    \includegraphics[width=\linewidth]{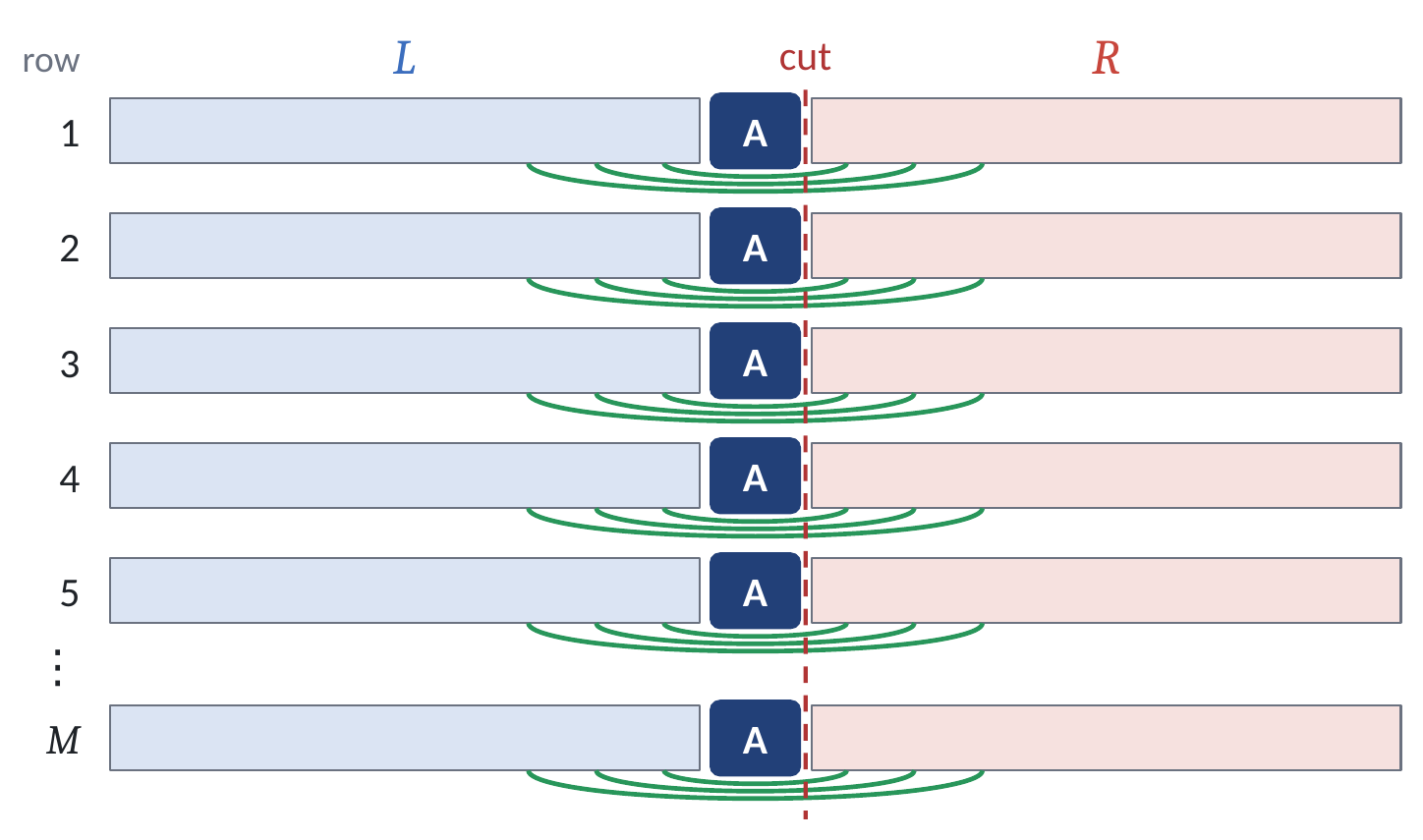}
  \end{minipage}
  \caption{\textbf{Local prefix activation and the projected parallel Bell sector.}
  (a) The central flags form $\mathsf A^t\mathsf D^{M-t}$, here with
  $t=3$. Shaded bands are active intervals. Only the prefix boundary can
  advance or retreat: the next empty row may activate, while the last
  active row may deactivate only after returning to its empty root.
  The local rule acts on at most three neighboring flags and two central
  row sites, as in Eq.~\eqref{app:eq:activation-rule}.
  The one-site well $-\Pi_{\mathsf D_1}$ singles out the global product
  root in the legal subspace.
  (b) After $P_F^{(M)}$ selects all flags active and every row in its
  target, the normalized state contains a common factor of $Mm$ Bell
  pairs. The cut lies immediately to the right of the flags, which
  belong to $L$. Green arcs depict this conditional factor, whose weight
  in the full ground state is $w_{M,m}$, not one.
  The fully active diagonal block is $K_M+W_M$, with
  $W_M=\tfrac12\Pi_a^{(M)}\le\tfrac12 I$; it is not a sum of
  independent rooted row Hamiltonians. A single root and common
  resolvent yield the startup and target-weight estimates in
  Eqs.~\eqref{app:eq:startup-plus-transport} and
  \eqref{app:eq:parallel-weight}.}
  \label{fig:app-parallel}
\end{figure}

Figure~\ref{fig:app-parallel} illustrates the local activation rule and the conditional Bell sector.

For row $j=1,\ldots,M$, put its left sites at $x=-b,\ldots,-1$ and its
right sites at $x=1,\ldots,b$. At $(0,j)$ place a flag with states
$\mathsf A_j$ (active) and $\mathsf D_j$ (dormant). The flag space can
have dimension two; it can also be embedded in six dimensions with a
one-site penalty on unused states. Use Manhattan distance.

Impose the prefix grammar
\begin{align}
 \mathsf A^t\mathsf D^{M-t},\qquad t=0,\ldots,M,
 \label{app:eq:prefix-grammar}
\end{align}
by forbidding adjacent $\mathsf D\mathsf A$. Also impose the local
constraint that a dormant flag has $UU$ on its two central row sites.
Equivalently, add the diagonal projector
$\Pi_{\mathsf D_j}(I-\Pi_{UU}^{(j)})$.
The row grammar then forces a dormant row to be entirely $U$.
Control only its start term by $\Pi_{\mathsf A_j}$; all other row
propagation terms are unchanged and automatically annihilate the
all-$U$ dormant row. No distant bond is controlled by a nonlocal flag
condition.

Fix virtual endpoint flags $\mathsf A_0$ and $\mathsf D_{M+1}$.
The activation projector is
\begin{align}
 h_j^{\mathrm{act}}=
 \Pi_{\mathsf A_{j-1}}\Pi_{\mathsf D_{j+1}}\Pi_{UU}^{(j)}
 \frac{(|\mathsf D_j\rangle-|\mathsf A_j\rangle)
       (\langle\mathsf D_j|-\langle\mathsf A_j|)}2.
 \label{app:eq:activation-rule}
\end{align}
Projectors onto the virtual flags are interpreted as identities.
This term involves at most three flags and two central row sites, so it
is at most five-local and has support diameter at most two.
It activates an empty next row, or deactivates the last active row only
when that row has returned to its root.

Let $\widehat J_{\mathrm{phys}}\ge0$ be the sum of these activation
terms and the row propagation terms, including the controlled starts.
Let $\widehat\Gamma$ be the sum of the row grammar, prefix grammar,
dormancy, and any unused-flag-state penalties. Every propagation and
activation term preserves the global legal subspace: row terms vanish
on dormant rows and preserve an active row's grammar; activation flips
only an empty row at the prefix boundary. Hermiticity also makes the
illegal subspace invariant. Define
\begin{align}
 H_{M,b}^{(2),\mathrm{raw}}
 =\widehat J_{\mathrm{phys}}+2\widehat\Gamma-\Pi_{\mathsf D_1}.
 \label{app:eq:H-parallel}
\end{align}
Inside the legal subspace, the first flag being dormant forces every
flag to be dormant and every row to be all $U$. Thus
$\Pi_{\mathsf D_1}$ is exactly the projector onto the single product root
\begin{align}
 |o\rangle=\bigotimes_{j=1}^M
           \bigl(|\mathsf D_j\rangle\otimes|U\rangle_j^{\otimes2b}\bigr).
 \label{app:eq:global-root}
\end{align}
Only the first activation term contributes at this root, and
$\langle o|\widehat J_{\mathrm{phys}}|o\rangle=1/2$.
Lemma~\ref{app:lem:well} implies uniqueness, $1/2\le\lambda_*\le1$,
and an unscaled full-system gap of at least $1/2$.
The illegal subspace has energy at least one.

There are $n=M(2b+1)$ sites. Put all flags in $L$, with the cut between
$x=0$ and $x=1$, so $\mathcal A=M$ exactly. The norms and the number of
terms at every site are uniformly bounded. Enlarge the fixed constant
$g_*$ from the row construction if necessary and set
$H_{M,b}^{(2)}=H_{M,b}^{(2),\mathrm{raw}}/g_*$.
This gives the local-strength normalization and the uniform physical gap
$\Delta_0=1/(2g_*)$, without changing the ground state.

\subsection{The reduced prefix graph and a constant exit cost}
\label{app:parallel-blocks}

The initialized history space is isometrically identified with
\begin{align}
 \widehat{\mathcal H}_{\mathrm{hist}}
 \cong\bigoplus_{t=0}^M\bigl(\mathbb C^{\mathcal Q_b}\bigr)^{\otimes t}.
 \label{app:eq:prefix-history}
\end{align}
The fixed dormant rows and flags are suppressed; the $t=0$ summand is the
one-dimensional root space. The row isometry of
Lemma~\ref{app:lem:history} identifies each active factor. Activation
adjoins $|a\rangle$ and deactivation contracts with $\langle a|$.
Consequently this history space is invariant and contains $o$, so
Lemma~\ref{app:lem:well} puts the ground state inside it.

Write $\widehat J$ for the restriction of
$\widehat J_{\mathrm{phys}}$ in the history basis. Let
$V_t\psi=\psi\otimes|a\rangle$. The off-diagonal block from $t$ to
$t+1$ is $-V_t/2$ and its reverse is $-V_t^\dagger/2$.
The diagonal block is
\begin{align}
 K_t+W_t,\qquad
 K_t:=\sum_{j=1}^tJ_b^{(j)},\qquad
 W_t:=\tfrac12\mathbf1_{t<M}I
           +\tfrac12\mathbf1_{t>0}\Pi_a^{(t)},
 \label{app:eq:prefix-blocks}
\end{align}
where $K_0=0$, and for $t\ge1$,
$\Pi_a^{(t)}=I^{\otimes(t-1)}\otimes|a\rangle\langle a|$.
The second term in $W_0$ is absent. Thus
\begin{align}
 0\le W_t\le I,\qquad
 W_M=\tfrac12\Pi_a^{(M)}\le\tfrac12 I.
 \label{app:eq:exit-cost}
\end{align}
There is only one deactivation channel at the end of the active prefix,
not one channel for every row. This is the reason the final-sector exit
cost is independent of $M$.

\subsection{A linear startup cost}
\label{app:parallel-startup}

All matrix inequalities in this subsection are \emph{entrywise in the
reduced history basis}. We do not use operator-monotonicity of the
exponential in the Loewner order, or stoquasticity on the full physical
computational basis. In the history basis, $J_b$, every $K_t$, and
$\widehat J$ have nonpositive off-diagonal entries. Their heat kernels
are therefore entrywise nonnegative: shift by a sufficiently large
scalar and expand the exponential of an entrywise nonnegative matrix.

The row's positive zero eigenvector is $\varphi_\pi=\sqrt\pi$.
The spectral expansion gives
\begin{align}
 \langle a|e^{-sJ_b}|a\rangle\ge
      |\langle a|\varphi_\pi\rangle|^2=p\ge\tfrac13
      \qquad(s\ge0).
 \label{app:eq:return-kernel}
\end{align}
Since $W_t$ is diagonal and $W_t\le I$, the Trotter product formula
also gives entrywise
\begin{align}
 e^{-s(K_t+W_t)}\ge e^{-s}e^{-sK_t}.
 \label{app:eq:killed-comparison}
\end{align}
Indeed, $e^{-sW_t/N}\ge e^{-s/N}I$ entrywise, and all the other factors
in the product have nonnegative entries.

\begin{applemma}[Startup amplitude]
\label{app:lem:startup}
Let $a^M$ denote the product row source in the fully active sector.
For every $t_0>0$,
\begin{align}
 \langle a^M|e^{-t_0\widehat J}|o\rangle
 \ge e^{-t_0}\left(\frac p2\right)^M\frac{t_0^M}{M!}.
 \label{app:eq:startup-amplitude}
\end{align}
In particular, at $t_0=M$ this is at least $e^{-C_{\mathrm{act}}M}$,
where $C_{\mathrm{act}}=1+\log(2/p)\le1+\log6$.
\end{applemma}
\begin{appproof}
Use the Dyson expansion around the block-diagonal part
$\bigoplus_t(K_t+W_t)$. Its interaction term has nonnegative off-diagonal
blocks in the expansion of $e^{-t_0\widehat J}$.
Retain only trajectories with exactly the $M$ forward activations
$0\to1\to\cdots\to M$ at times $0<t_1<\cdots<t_M<t_0$.
Each contributes a factor $1/2$. Equation~\eqref{app:eq:killed-comparison}
bounds the combined diagonal activation cost below by $e^{-t_0}$.
After its birth at $t_j$, row $j$ evolves independently for time
$t_0-t_j$; its final return amplitude is at least $p$ by
\eqref{app:eq:return-kernel}. Their product is at least $p^M$.
Integrating over the activation simplex gives its volume $t_0^M/M!$.
At $t_0=M$, use $M^M/M!\ge1$.
Importantly, rows are allowed to move during startup; they are not
required to remain frozen at the source.
\end{appproof}

Define the normalized row target witness and its kernel by
\begin{align}
 |v_F\rangle:=q_F^{-1/2}\sum_{w\in F_m}\sqrt{\pi(w)}|w\rangle,
 \qquad k_m(u):=\langle v_F|e^{-uJ_b}|a\rangle.
 \label{app:eq:target-kernel}
\end{align}
In the fully active sector take
$|\widehat v_F\rangle=|v_F\rangle^{\otimes M}$.
In the semigroup composition at time $M$, retain the intermediate state
$a^M$ and subsequently retain only propagation within the fully active
sector. By \eqref{app:eq:exit-cost}, its heat kernel is entrywise at least
$e^{-u/2}e^{-uK_M}$. Lemma~\ref{app:lem:startup} then yields
\begin{align}
 \langle\widehat v_F|e^{-(M+u)\widehat J}|o\rangle
 \ge e^{-C_{\mathrm{act}}M-u/2}\,k_m(u)^M,
 \qquad u\ge0.
 \label{app:eq:startup-plus-transport}
\end{align}
Thus the startup time is $M$, its logarithmic cost is $O(M)$, and the
additional killing rate is $1/2$. No Poincar\'e estimate for the full
prefix graph is needed.

\subsection{From a Poincar\'e constant to a resolvent lower bound}
\label{app:parallel-resolvent}

\begin{applemma}[Poincar\'e--resolvent inequality]
\label{app:lem:resolvent}
Let $J$ be a real symmetric matrix on a finite set of at least two
vertices, with $J_{xy}\le0$ for $x\ne y$ and connected off-diagonal
graph. Suppose $\pi(x)>0$, $\sum_x\pi(x)=1$, and $J\sqrt\pi=0$.
Let $\tau$ be the best Poincar\'e constant of the associated Doob chain.
For a source $a$ and target $F$, write $p=\pi(a)>0$, $q=\pi(F)>0$, and
$|v_F\rangle=q^{-1/2}\sum_{x\in F}\sqrt{\pi(x)}|x\rangle$.
For every $s>0$,
\begin{align}
 s\langle v_F|(J+sI)^{-1}|a\rangle
 \ge\sqrt{pq}\,
    \exp\!\left[-(p^{-1/2}+q^{-1/2})\sqrt{s\tau}\right].
 \label{app:eq:resolvent-lemma}
\end{align}
\end{applemma}
\begin{appproof}
Set $G=\operatorname{diag}(\sqrt\pi)$ and
$\mathsf L=G^{-1}JG$. With
$q_{xy}=-J_{xy}\sqrt{\pi(y)/\pi(x)}\ge0$ for $x\ne y$, we have
\begin{align}
 (\mathsf Lf)(x)=\sum_{y\ne x}q_{xy}(f(x)-f(y)),\qquad
 c_{xy}=\pi(x)q_{xy}=c_{yx}.
 \label{app:eq:Doob-transform}
\end{align}
Thus $\mathsf L$ is the nonnegative-sign Markov Laplacian, whose
transition semigroup is $e^{-t\mathsf L}$, and
$\langle f,\mathsf Lg\rangle_\pi=\mathcal E(f,g)$.
In particular $J\ge0$. Define
\begin{align}
 h:=s(\mathsf L+sI)^{-1}\frac{\mathbf1_a}{p}.
 \label{app:eq:resolvent-density}
\end{align}
Connectivity and the positive resolvent kernel give $h>0$.
Stationarity gives $\mathbb E_\pi h=1$. At a positive maximum away from
$a$, $(\mathsf L+sI)h>0$ would contradict the resolvent equation, so
$h(a)=\max h\ge1$.

Take the $L^2(\pi)$ inner product of
$(\mathsf L+sI)h=s\mathbf1_a/p$ with $1/h$. Since
$\langle1/h,h\rangle_\pi=1$, this gives
\begin{align}
 -\mathcal E(1/h,h)=s\left(1-\frac1{h(a)}\right)\le s.
 \label{app:eq:log-energy-start}
\end{align}
For $x,y>0$, Cauchy--Schwarz applied to $\int_y^x dt/t$ shows
$(\log x-\log y)^2\le(x-y)^2/(xy)$. Hence
\begin{align}
 \mathcal E(\log h,\log h)
 \le-\mathcal E(1/h,h)\le s,
 \qquad \operatorname{Var}_\pi(\log h)\le s\tau.
 \label{app:eq:log-density-variance}
\end{align}
Let $\ell_h=\mathbb E_\pi\log h$. The source contribution to the variance
and $\log h(a)\ge0$ imply $\ell_h\ge-\sqrt{s\tau/p}$.
A second Cauchy--Schwarz bound gives
$|\mathbb E_{\pi(\cdot|F)}\log h-\ell_h|\le\sqrt{s\tau/q}$.
Therefore
\begin{align}
 \mathbb E_{\pi(\cdot|F)}\log h
 \ge-(p^{-1/2}+q^{-1/2})\sqrt{s\tau}.
 \label{app:eq:target-log-density}
\end{align}
Finally, the Doob transform gives the exact identity
\begin{align}
 s\langle v_F|(J+sI)^{-1}|a\rangle
       =\sqrt{pq}\,\mathbb E_{\pi(\cdot|F)}h.
 \label{app:eq:resolvent-density-identity}
\end{align}
Jensen's inequality completes the proof.
\end{appproof}

For the single row, \eqref{app:eq:root-mass},
\eqref{app:eq:target-mass}, and Lemma~\ref{app:lem:poincare} all hold for
the same $J_b$ with $b=16m$. Since
$(J_b+sI)^{-1}=\int_0^\infty e^{-su}e^{-uJ_b}\,du$, we obtain
\begin{align}
 B_m(s):=s\int_0^\infty e^{-su}k_m(u)\,du
 \ge c_0 e^{-C_0\sqrt s\,m^{3/2}},\qquad s>0,
 \label{app:eq:row-Laplace-bound}
\end{align}
where $c_0,C_0$ are fixed. Because $J_b\ge0$ and its heat kernel is
entrywise nonnegative, $0\le k_m(u)\le1$.

\subsection{The ground-state resolvent and a common imaginary time}
\label{app:parallel-weight}

On the initialized history subspace, the unscaled physical Hamiltonian
has the form $H=J-|o\rangle\langle o|$, with $(J,o)=(J_b,a)$ in one
dimension and $(\widehat J,o)$ in the parallel construction. This is a
restriction to the history space, not an identity on the full physical
space with its grammar penalties. The eigenvalue equation at energy
$-\lambda_*<0$ gives
\begin{align}
 |\Omega\rangle=
 \frac{(J+\lambda_*I)^{-1}|o\rangle}
      {\|(J+\lambda_*I)^{-1}|o\rangle\|},
 \qquad \tfrac12\le\lambda_*\le1,
 \label{app:eq:ground-resolvent}
\end{align}
with positive root phase. Since $J\ge0$, the denominator is at most
$1/\lambda_*$. For a normalized target witness $v$ with $P_Fv=v$,
\begin{align}
 \|P_F\Omega\|^2
 \ge\left|\lambda_*\langle v|(J+\lambda_*I)^{-1}|o\rangle\right|^2.
 \label{app:eq:weight-from-resolvent}
\end{align}
Taking $v=v_F$, $o=a$, and $s=\lambda_*$ in
\eqref{app:eq:row-Laplace-bound} proves the one-dimensional estimate
\begin{align}
 w_m\ge e^{-Cm^{3/2}}.
 \label{app:eq:one-row-weight}
\end{align}

For the parallel construction, define for $\kappa>0$
\begin{align}
 A_{M,m}(\kappa):=\kappa\int_0^\infty
                    e^{-\kappa u}k_m(u)^M\,du.
 \label{app:eq:common-time-integral}
\end{align}
When $M\ge2$ and $s>\kappa/M$, H\"older's inequality gives
\begin{align}
 \frac{B_m(s)}s
 \le\left(\frac{A_{M,m}(\kappa)}\kappa\right)^{1/M}
    \left(\frac{M-1}{Ms-\kappa}\right)^{(M-1)/M}.
 \label{app:eq:holder-step}
\end{align}
This follows by factoring the integrand as
$(e^{-\kappa u/M}k_m(u))e^{-(s-\kappa/M)u}$.
At $s=2\kappa/M$, rearranging yields the exact prefactor and then the
simpler lower bound
\begin{align}
 A_{M,m}(\kappa)
 &\ge\frac{M^M}{2^M(M-1)^{M-1}}B_m(2\kappa/M)^M
 \ge2^{-M}B_m(2\kappa/M)^M\notag\\
 &\ge\exp[-C_1M-C_2\sqrt{M\kappa}\,m^{3/2}].
 \label{app:eq:common-time-bound}
\end{align}
For $M=1$, use $A_{1,m}(\kappa)=B_m(\kappa)$ directly.

Insert \eqref{app:eq:startup-plus-transport} into the resolvent integral,
retaining only times $t=M+u$, and put $\kappa=\lambda_*+1/2$.
Then
\begin{align}
 \lambda_*\langle\widehat v_F|
            (\widehat J+\lambda_*I)^{-1}|o\rangle
 &\ge e^{-(C_{\mathrm{act}}+\lambda_*)M}
       \frac{\lambda_*}{\lambda_*+1/2}
       A_{M,m}(\lambda_*+1/2)\notag\\
 &\ge\exp\!\left[-C\left(M+\sqrt M\,m^{3/2}\right)\right].
 \label{app:eq:parallel-resolvent-bound}
\end{align}
Here $1/2\le\lambda_*\le1$, so the common killing parameter is bounded
above and below independently of $M$. Squaring via
\eqref{app:eq:weight-from-resolvent} and absorbing constants proves
\begin{align}
 w_{M,m}\ge
 \exp\!\left[-C\left(M+\sqrt M\,m^{3/2}\right)\right].
 \label{app:eq:parallel-weight}
\end{align}
The common-time integral is essential: the construction has one global
negative well, not independent negative wells on the separate rows.

The target requires all flags to be active and at least $m$ particles
in the left half of each row. Since all flags belong to $L$, this is
recognized by one fixed $P_F=P_L\otimes I_R$.
Applying the row isometries from Lemma~\ref{app:lem:history} coherently
in every row gives the common $Mm$-Bell-pair factor in
\eqref{app:eq:bell-theorem}, even for correlated superpositions of row
configurations. This proves the physical construction, gap, and target
weight parts of Theorem~\ref{app:thm:lower} in one and two dimensions.

\FloatBarrier
\section{Higher-dimensional embedding and rank and entropy consequences}
\label{app:consequences}

\subsection{A cut-preserving serpentine embedding in any fixed dimension}
\label{app:embedding}

Let $\Sigma=\prod_{j=2}^{\nu}\{1,\ldots,N_j\}$ be a rectangular
cross-section, with $M=|\Sigma|$. It has a nearest-neighbor Hamiltonian
path $\gamma(1),\ldots,\gamma(M)$, meaning a path that visits each
cross-sectional site once. One explicit construction is inductive:
traverse a one-dimensional interval in order, and traverse successive
slices of a higher-dimensional box using this order and its reverse
alternately. The last vertex of one slice and first vertex of the next
then differ by one step in the new coordinate. This also covers side
lengths equal to one.

Relabel every site of the two-dimensional construction by
\begin{align}
 (x,j)\longmapsto(x,\gamma(j)),\qquad -b\le x\le b.
 \label{app:eq:serpentine-map}
\end{align}
Equivalently, if $\mathcal U_\gamma$ is the induced permutation of tensor
factors, define
\begin{align}
 H_{M,b}^{(\nu)}
 :=\mathcal U_\gamma H_{M,b}^{(2)}\mathcal U_\gamma^\dagger.
 \label{app:eq:embedded-H}
\end{align}
The row terms still act along $x$. A pair of neighboring flags lies at
neighboring sites of $\Sigma$. The largest activation term is supported
on
\begin{align}
 \{(0,\gamma(j-1)),(0,\gamma(j)),(0,\gamma(j+1)),
                 (-1,\gamma(j)),(1,\gamma(j))\},
 \label{app:eq:activation-support}
\end{align}
with nonexistent endpoint flags omitted. Every listed site lies within
Manhattan distance one of $(0,\gamma(j))$, so the support diameter is
at most two, including at turns of the serpentine path. The dormant-row
constraints and the controlled starts obey the same diameter bound.
Thus $d_0\le6$, $k\le5$, and $r_0\le2$ remain unchanged.
Each site's incident terms are precisely relabeled terms of the
original construction, so the same strength normalization $g_*$ works.

The relabeling never moves a site from $L$ to $R$, or conversely:
\begin{align}
 \mathcal U_\gamma=
      \mathcal U_{\gamma,L}\otimes\mathcal U_{\gamma,R}.
 \label{app:eq:cut-factorized-permutation}
\end{align}
It therefore preserves the full spectrum, uniqueness, the gap, the root
overlap, every Schmidt coefficient, the product character of $P_F$, and
its success probability. Bell extraction is transported by the same
local isometries. The result is a Hamiltonian on the full rectangular
box with $n=M(2b+1)$ and $\mathcal A=M$ exactly. No interactions on the
unused transverse bonds are required. This proves the geometric part
of Theorem~\ref{app:thm:lower} for every fixed $\nu\ge2$.

\subsection{Transferring a fixed Bell sector to the Schmidt tail}
\label{app:tail-transfer}

Write $w=\|P_F\Omega\|^2>0$ and $d=2^{Mm}$. By
\eqref{app:eq:bell-theorem}, the normalized target has Schmidt
probabilities equal to the Schmidt probabilities of $\Xi$, each divided
by $d$ and repeated $d$ times. In particular, every target Schmidt
probability is at most $1/d$.

For any possibly unnormalized vector $\phi$ of Schmidt rank at most
$D$, product projection is contractive and cannot increase its rank.
The best rank-$D$ approximation to $P_F\Omega$ is its Schmidt
truncation, and it retains squared norm at most $Dw/d$. Therefore
\begin{align}
 \|\Omega-\phi\|^2
 \ge\|P_F\Omega-P_F\phi\|^2
 \ge w\left(1-\frac Dd\right),\qquad 1\le D\le d.
 \label{app:eq:tail-transfer}
\end{align}
This proves \eqref{app:eq:finite-tail-theorem} and completes
Theorem~\ref{app:thm:lower}. In particular,
\begin{align}
 \inf_{\operatorname{SR}(\phi)\le2^{Mm-1}}\|\Omega-\phi\|^2
 \ge\tfrac12\exp\!\left[-C\left(M+\sqrt M\,m^{3/2}\right)\right].
 \label{app:eq:half-rank-tail}
\end{align}
Possible entanglement in the residual state $\Xi$ cannot weaken this
bound. More generally, if the normalized target has Schmidt probabilities
$\{\mu_j^2\}$, the same argument gives
$\sum_{j>D}\lambda_j^2\ge w\sum_{j>D}\mu_j^2$ for the full state's
Schmidt probabilities $\{\lambda_j^2\}$. This is the fixed-product-projection tail transfer used in the error-space discussion of the main
text.

\subsection{Choosing the accuracy and system size}
\label{app:accuracy-proof}

Let $C$ be the constant in \eqref{app:eq:half-rank-tail}. First choose a
fixed $c>0$ so small that $Cc^{3/2}\le1/2$, and then choose
\begin{align}
 C_0\ge\max\{2C,\ 2\log100+\log2+1,\ (4/c)^{3/2}\}.
 \label{app:eq:accuracy-constants}
\end{align}
For $L_\epsilon\ge C_0M$, take
\begin{align}
 m=\left\lfloor c\frac{L_\epsilon^{2/3}}{M^{1/3}}\right\rfloor.
 \label{app:eq:chosen-m}
\end{align}
The quantity inside the floor is at least four, so $m\ge3$ and
$m=\Theta(L_\epsilon^{2/3}M^{-1/3})$ uniformly. Moreover,
\begin{align}
 CM+C\sqrt M\,m^{3/2}\le L_\epsilon.
 \label{app:eq:chosen-exponent}
\end{align}
Since $\epsilon^2=10^4e^{-2L_\epsilon}$, our explicit constant choice
ensures
\begin{align}
 \tfrac12 e^{-L_\epsilon}>10^4e^{-2L_\epsilon}=\epsilon^2.
 \label{app:eq:accuracy-comparison}
\end{align}
Equation~\eqref{app:eq:half-rank-tail} thus rules out every approximation
of Schmidt rank at most $2^{Mm-1}$ at error $\epsilon$, and
\begin{align}
 \log D_\epsilon(\Omega)
 \ge c'Mm\ge c''(ML_\epsilon)^{2/3}.
 \label{app:eq:accuracy-rank-conclusion}
\end{align}
With $\mathcal A=M$, this is \eqref{app:eq:state-accuracy-theorem}.
The constant shift between $L_\epsilon=\log(100/\epsilon)$ and
$\log(1/\epsilon)$ has been explicitly absorbed, rather than dropped at
finite accuracy. In one dimension the same argument uses
\eqref{app:eq:one-row-weight} and $m=\Theta(L_\epsilon^{2/3})$.

\paragraph{Approximate AGSPs.}
For the definition in \eqref{app:eq:agsp-definition}, Hermiticity and
$K\Omega_K=\Omega_K$ give
\begin{align}
 \|K-|\Omega\rangle\langle\Omega|\|
 \le\|K-|\Omega_K\rangle\langle\Omega_K|\|
       +\||\Omega_K\rangle\langle\Omega_K|
                          -|\Omega\rangle\langle\Omega|\|
 \le\epsilon_K+\delta_K=\eta_K.
 \label{app:eq:AGSP-projector-distance}
\end{align}
Choose the phase so that $a_o:=\langle\Omega|o\rangle>0$.
Lemma~\ref{app:lem:well} gives $a_o^2\ge1/2$, and $o$ is a product state.
Consequently
\begin{align}
 \left\|\frac{K|o\rangle}{a_o}-|\Omega\rangle\right\|
      \le\sqrt2\eta_K,
 \qquad
 \operatorname{SR}(K|o\rangle)\le\operatorname{SR}(K)\le D_K.
 \label{app:eq:AGSP-to-state}
\end{align}
The rescaling by $a_o$ is allowed because approximate Schmidt rank does
not require normalized approximants.

To make the quantifiers precise, prescribe $M$ and $0<\eta\le1/8$
first, with $\log(1/\eta)\ge C_1M$ for a sufficiently large fixed $C_1$.
Put $\epsilon'=\sqrt2\eta$ and
$L_{\epsilon'}=\log(100/\epsilon')$.
Then $\epsilon'\le1/4$ and $L_{\epsilon'}\ge C_0M$.
Choose one Hamiltonian using \eqref{app:eq:chosen-m} with this
$L_{\epsilon'}$. For every AGSP on that Hamiltonian satisfying
$\eta_K\le\eta$, \eqref{app:eq:AGSP-to-state} and
\eqref{app:eq:accuracy-rank-conclusion} imply
\begin{align}
 \log D_K\ge c_{\mathrm{agsp}}
              \left(\mathcal A\log\frac1\eta\right)^{2/3}.
 \label{app:eq:AGSP-quantified}
\end{align}
The same proof applies to the one-dimensional product root with
$\mathcal A=1$. This proves Corollary~\ref{app:cor:accuracy}.

Once this finite system has been selected, \eqref{app:eq:AGSP-quantified}
does not replace its prescribed $\eta$ by an arbitrarily smaller actual
error $\eta_K$. In particular, the exact projector has finite Schmidt
rank on every fixed finite system. Result~\ref{result:main-lower} in the main text uses a common prescribed
accuracy $\epsilon$ for the state and AGSP bounds. To obtain both on the
same Hamiltonian, set $\eta=\epsilon$ and choose the construction for
$\epsilon'=\sqrt2\epsilon$ as above. Since
$D_\epsilon(\Omega)\ge D_{\sqrt2\epsilon}(\Omega)$ and
$L_{\sqrt2\epsilon}=L_\epsilon-\tfrac12\log2$, both lower bounds follow
after adjusting the fixed constants. At fixed positive gap and in the
high-accuracy regime, this quantified lower bound and
Theorem~\ref{app:thm:upper} match in their $2/3$ exponent and joint
boundary--accuracy dependence, up to logarithmic factors.

\subsection{R\'enyi entropy and the hypercubic volume law}
\label{app:renyi-proof}

\begin{appcor}[R\'enyi volume laws]
\label{app:cor:renyi}
For the states in Theorem~\ref{app:thm:lower} and every $0<\alpha<1$,
\begin{align}
 S_\alpha(\Omega_{M,m}^{(\nu)})
 \ge Mm\log2-C\frac{\alpha}{1-\alpha}
                   \left(M+\sqrt M\,m^{3/2}\right).
 \label{app:eq:renyi-theorem}
\end{align}
In two dimensions, for each fixed $\alpha$, taking
$m=\lfloor a_\alpha M\rfloor$ with a sufficiently small
$a_\alpha>0$ yields $S_\alpha\ge c_\alpha n$ for some
$c_\alpha>0$ independent of $n$, once the system is sufficiently large
at that fixed $\alpha$.
For every fixed $\nu\ge3$, take $R=32m+1$, $M=R^{\nu-1}$, and
$n=R^\nu$. This single hypercubic family, independent of $\alpha$,
satisfies
\begin{align}
 S_\alpha(\Omega_R^{(\nu)})
 \ge\frac{\log2}{32}R^\nu
       -O_\alpha\!\left(R^{\nu-1}+R^{(\nu+2)/2}\right),
 \qquad
 \liminf_{\substack{R\to\infty\\R=32m+1}}
 \frac{S_\alpha(\Omega_R^{(\nu)})}{R^\nu}
 \ge\frac{\log2}{32}.
 \label{app:eq:hypercube-theorem}
\end{align}
These conclusions hold for each fixed $\alpha<1$, not uniformly as
$\alpha\uparrow1$. Appendix~\ref{app:adiabatic} proves a von Neumann area law for the same states.
\end{appcor}

\begin{appproof}[Proof of Corollary~\ref{app:cor:renyi}]
Regard a pure state as its coefficient matrix across $L|R$. Product
projection acts by contractions on its two sides. Every singular value
of the projected matrix is at most the corresponding singular value of
the original matrix. This follows, for example, by applying the
best-rank-$(j-1)$ operator-norm approximation characterization of the
$j$th singular value to $P_LXP_R^{\mathsf T}$ and any rank-$(j-1)$
approximation of $X$.
For $0<\alpha<1$, the Bell factor of dimension $d=2^{Mm}$ therefore
implies
\begin{align}
 \operatorname{tr}\rho_L^\alpha
 \ge w_{M,m}^{\alpha}d^{1-\alpha}
                    \operatorname{tr}\rho_{\Xi,L}^{\alpha}
 \ge w_{M,m}^{\alpha}d^{1-\alpha}.
 \label{app:eq:renyi-projection}
\end{align}
The last step uses normalization of $\Xi$: its eigenvalues lie in
$[0,1]$, and $x^\alpha\ge x$ there.
Taking logarithms and applying \eqref{app:eq:parallel-weight} gives
\begin{align}
 S_\alpha(\Omega_{M,m}^{(\nu)})
 \ge Mm\log2+\frac{\alpha}{1-\alpha}\log w_{M,m}
 \ge Mm\log2-C\frac{\alpha}{1-\alpha}
                  (M+\sqrt M\,m^{3/2}).
 \label{app:eq:renyi-finite-proof}
\end{align}

\paragraph{Two dimensions.}
Fix $0<\alpha<1$ and choose $a_\alpha>0$ such that
\begin{align}
 C\frac{\alpha}{1-\alpha}\sqrt{a_\alpha}\le\frac{\log2}{4}.
 \label{app:eq:aspect-choice}
\end{align}
For $m=\lfloor a_\alpha M\rfloor$ and sufficiently large $M$, the
transport term in \eqref{app:eq:renyi-finite-proof} is at most
$Mm\log2/4$, since $\sqrt M\,m^{3/2}=Mm\sqrt{m/M}$.
The startup term, proportional to $M$, is also at most $Mm\log2/4$
for all sufficiently large $M$. Thus
\begin{align}
 S_\alpha(\Omega_{M,m}^{(2)})\ge\tfrac12 Mm\log2
       \ge c_\alpha n,\qquad n=M(32m+1),
 \label{app:eq:2d-volume-proof}
\end{align}
with a positive coefficient $c_\alpha$ independent of $n$.
Only the rectangular aspect ratio depends on $\alpha$; the alphabet,
interaction range, strength normalization, and gap bound do not.

\paragraph{Dimensions three and above.}
Fix $\nu\ge3$, set $R=32m+1$, and take the cubic cross-section
$\Sigma=\{1,\ldots,R\}^{\nu-1}$. The embedding above gives a hypercube
with $M=R^{\nu-1}$ and $n=R^\nu$. Exactly,
\begin{align}
 Mm\log2=\frac{\log2}{32}(R^\nu-R^{\nu-1}),\qquad
 M=R^{\nu-1},\qquad
 \sqrt M\,m^{3/2}=O(R^{(\nu+2)/2}).
 \label{app:eq:hypercube-scaling}
\end{align}
For every fixed $0<\alpha<1$, substituting these expressions into
\eqref{app:eq:renyi-finite-proof} gives
\begin{align}
 S_\alpha(\Omega_R^{(\nu)})
 \ge\frac{\log2}{32}R^\nu
       -O_\alpha(R^{\nu-1}+R^{(\nu+2)/2}).
 \label{app:eq:hypercube-finish}
\end{align}
Both error exponents are strictly smaller than $\nu$ precisely when
$\nu>2$. Dividing by $R^\nu$ and taking the lower limit along
$R=32m+1$ proves \eqref{app:eq:hypercube-theorem}.
The Hamiltonian family has no $\alpha$-dependent geometric choice.
The correction nevertheless contains $\alpha/(1-\alpha)$, so its
convergence is not uniform as $\alpha\uparrow1$.
The same states obey a von Neumann area law by the independent gapped-path argument of Appendix~\ref{app:adiabatic}.
\end{appproof}

\clearpage
\subsection{The state-approximation consequence of the upper bound}
\label{app:seed-approximation}

For completeness, the AGSP upper bound also justifies the
constant-overlap-seed argument in the discussion of the main text.
Let $S_1=-\sum_j\lambda_j^2\log\lambda_j^2$ be the von Neumann
entanglement entropy across the chosen cut, with Schmidt coefficients
in decreasing order. Since $\lambda_j^2\le1/j$, taking
$\chi_0=\lceil e^{2S_1}\rceil$ gives
\begin{align}
 \sum_{j>\chi_0}\lambda_j^2
 \le\frac{S_1}{\log(\chi_0+1)}\le\tfrac12.
 \label{app:eq:seed-tail}
\end{align}
The normalized Schmidt truncation $\phi_0$ therefore has rank at most
$\chi_0$ and positive overlap
$a_0=\langle\Omega|\phi_0\rangle\ge1/\sqrt2$.
Use Theorem~\ref{app:thm:upper} with operator error
$\eta=\epsilon/(4\sqrt2)$, where $0<\epsilon<1$.
For $v=K\phi_0$, we have $\|v-a_0\Omega\|\le\eta$, so $v\ne0$.
Normalization costs at most a second copy of this relative error:
\begin{align}
 \left\|\frac v{\|v\|}-\Omega\right\|
 \le\left|1-\frac{\|v\|}{a_0}\right|
       +\left\|\frac v{a_0}-\Omega\right\|
 \le\frac{2\eta}{a_0}\le\epsilon.
 \label{app:eq:seed-normalization}
\end{align}
Applying an operator of Schmidt rank $D_K$ to a vector of Schmidt rank
$\chi_0$ produces rank at most $D_K\chi_0$. At fixed positive gap,
$\log\chi_0\le2S_1+\log2$ and Theorem~\ref{app:thm:upper} imply
\begin{align}
 \log\operatorname{SR}\!\left(\frac v{\|v\|}\right)
 \le2S_1+\widetilde O\!\left(
       \mathcal A+[\mathcal A\log(e/\epsilon)]^{2/3}\right).
 \label{app:eq:seed-rank}
\end{align}
This is a one-cut existence statement. Constructing a global
tensor-network approximation requires an additional estimate on
accumulated errors across cuts, and finding such an approximation
efficiently is a separate algorithmic problem.

\clearpage
\section{A uniformly gapped path from a product state and the entropy separation}
\label{app:adiabatic}

\subsection{Overview and precise statement}

The counterexamples belong to a more restricted class than arbitrary
uniformly gapped ground states: each is connected to a product state by a
smooth local path whose gap and local interaction bounds are uniform in
system size. The path gives a von Neumann area law by quasi-adiabatic
continuation, but the same endpoint states have the R\'enyi volume laws
proved in Appendix~\ref{app:renyi-proof}. No general higher-dimensional
area-law theorem is needed for this separation.

\begin{appprop}[Uniformly gapped product-state path]
\label{app:prop:adiabatic-path}
For each lower-bound Hamiltonian constructed in
Appendices~\ref{app:row}--\ref{app:consequences}, write its unnormalized
physical operator as
\begin{align}
 H^{\rm raw}=J_{\rm phys}+2\Gamma-R_o.
 \label{app:eq:path-raw}
\end{align}
Here $J_{\rm phys}\ge0$ is the sum of propagation and, when present,
activation terms; $\Gamma$ is the sum of the local grammar penalties;
and $R_o$ is the local root-identifying projector. There is a fixed
normalization $g_*>0$ such that
\begin{align}
 H(s)=g_*^{-1}\bigl(sJ_{\rm phys}+2\Gamma-R_o\bigr),
 \qquad 0\le s\le1,
 \label{app:eq:adiabatic-path}
\end{align}
has a unique ground state $\Omega(s)$ for every $s$, with
\begin{align}
 \Omega(0)=o,\qquad \Omega(1)=\Omega,\qquad
 \Delta(s)\ge\frac{1-sr_o}{g_*}\ge\frac1{2g_*},
 \qquad r_o:=\langle o|J_{\rm phys}|o\rangle\le\tfrac12.
 \label{app:eq:path-gap}
\end{align}
The state $o$ is a product of on-site states. The paths have on-site
dimension at most six, interaction size at most five and range at most
two. Both their local interaction strength and the local strength of
$\partial_sH(s)$ are bounded uniformly. For the fixed planar cut,
\begin{align}
 S_1(\Omega(s))\le C_\nu\mathcal A,\qquad 0\le s\le1,
 \label{app:eq:path-area}
\end{align}
where $C_\nu$ is independent of $M,m,s$ and of the entropy index used to
select the endpoint family.
\end{appprop}

For one row, $R_o=\Pi_{UU}$, $o=a$, and $r_o=1/(b+1)$.
For the parallel construction, $R_o=\Pi_{D_1}$ and $r_o=1/2$.
The higher-dimensional construction is a tensor-factor permutation of
the latter model. Its terms retain the same support-size and range bounds, and the same
strengths, as established in Appendix~\ref{app:consequences}.

\subsection{The gap on the entire path}

\begin{appproof}[Proof of the path and gap assertions]
Every term preserves the legal subspace and its orthogonal complement.
On the legal subspace, $\Gamma=0$ and $R_o=|o\rangle\langle o|$.
Consequently, $g_*H(s)$ restricts to
$sJ_{\rm phys}-|o\rangle\langle o|$. A rank-one negative perturbation
of a positive operator has at most one negative eigenvalue. The root
trial state has expectation
\begin{align}
 \langle o|g_*H(s)|o\rangle=sr_o-1\le-\tfrac12.
 \label{app:eq:path-trial}
\end{align}
Thus a simple negative eigenvalue $-\lambda(s)$ exists, and
$1-sr_o\le\lambda(s)\le1$. All other legal eigenvalues are
nonnegative. On the illegal subspace,
\begin{align}
 sJ_{\rm phys}+2\Gamma-R_o\ge2I-I=I.
 \label{app:eq:path-illegal}
\end{align}
The ground state is therefore unique on the full physical space, and its
gap is at least $\lambda(s)/g_*$. At $s=0$ its ground vector is exactly
$o$. At $s=1$ it is the previously constructed state. This is the
rank-one argument of Appendix~\ref{app:row-gap}, applied for every $s$;
it never requires a lower bound on the internal propagation gap.

Multiplying positive propagation terms by $s\in[0,1]$ cannot increase
any local norm. The derivative is $J_{\rm phys}/g_*$, with the same
finite supports and uniformly bounded interaction strength. Increase
the one fixed $g_*$, if necessary, to bound both strengths by one.
Finite-dimensional spectral perturbation theory gives a smooth choice
of the unique ground-state projector along this linear path. The same
argument holds after serpentine re-embedding, since the embedding does
not depend on $s$.
\end{appproof}

\subsection{Why the von Neumann area law follows}

\begin{appproof}[Proof of \eqref{app:eq:path-area}]
We apply the entangling-rate and exact quasi-adiabatic-continuation
results of Refs.~\cite{PhysRevB.72.045141,PhysRevLett.111.170501,Marien2016}.
For exact spectral transport and its uniform quasi-locality we use the
finite-volume spectral-flow estimates of
Ref.~\cite{Bachmann2012Automorphic}, in particular Proposition~2.4 and
Theorem~4.8. The local propagation estimates are of Lieb--Robinson
type~\cite{ref:Nachtergaele2006-LR}.
Here it is important to check uniformity even though some microscopic
couplings depend on the chosen finite size. The path has a common gap
lower bound, common finite interaction range, bounded local dimension,
and uniform local bounds for both $H(s)$ and $\partial_sH(s)$.
These are precisely the inputs to a uniform Lieb--Robinson bound and
to a uniform exact spectral-flow filter. No translation invariance or
lower bound on each nonzero coupling is needed for the resulting
finite-volume estimates.

For clarity, the geometric summation underlying the application is as
follows. The ground states can be transported by a unitary generated
by a quasi-local Hermitian operator $\mathcal G(s)$ with a decomposition
\begin{align}
 \mathcal G(s)=\sum_{i\in\Lambda}\sum_{r\ge1}g_{i,r}(s),\qquad
 \operatorname{supp}g_{i,r}(s)\subseteq B(i,c r),\qquad
 \sup_s\|g_{i,r}(s)\|\le C_N(1+r)^{-N}.
 \label{app:eq:path-quasilocal}
\end{align}
For every fixed $N$, the constants $c,C_N$ depend only on the uniform
local data and gap, not on the volume. This is the standard
quasi-local decomposition of exact spectral flow. Terms supported
entirely on one side of the cut do not change its entanglement.
A radius-$cr$ ball contains $O(r^\nu)$ sites, and at most
$O(\mathcal A r)$ centers have such a ball crossing a planar cut.
The small-incremental-entangling bound (Theorem~3 of
Ref.~\cite{Marien2016}) controls each crossing term by
its norm times the logarithm of its supported Hilbert-space dimension,
independently of the remaining sites treated as ancillas. Hence the
integrated entangling-rate bound gives
\begin{align}
 |S_1(\Omega(s))-S_1(\Omega(0))|
 \le C\mathcal A\log6\int_0^s
       \sum_{r\ge1}r^{\nu+1}C_N(1+r)^{-N}\,du
 \le C_\nu\mathcal A
 \label{app:eq:path-rate-integral}
\end{align}
for any fixed $N>\nu+2$. The integrated form also covers points at
which a reduced density matrix has zero eigenvalues. Finally,
$S_1(\Omega(0))=0$ because $o$ is a product state.
This explicitly verifies the finite-size-uniform area-law stability
argument for the present rectangular geometries.
\end{appproof}

\begin{appcor}[Area/volume separation in an adiabatically connected family]
\label{app:cor:adiabatic-separation}
For each fixed $0<\alpha<1$ in two dimensions, the aspect-ratio choice
of Corollary~\ref{app:cor:renyi} gives, on the same family,
\begin{align}
 S_1(\Omega)=O(\mathcal A),\qquad
 S_\alpha(\Omega)\ge c_\alpha n.
 \label{app:eq:adiabatic-separation}
\end{align}
Here $c_\alpha>0$ is independent of $n$, and the inequalities hold
for sufficiently large systems at the fixed $\alpha$.
For every fixed $\nu\ge3$, one and the same hypercubic family has this
separation for every fixed $0<\alpha<1$. All these families have the
uniformly gapped product-state paths of
Proposition~\ref{app:prop:adiabatic-path}. In particular, their
$S_\alpha$ also obey an area law for $\alpha\ge1$ by monotonicity.
\end{appcor}

\begin{appproof}
Combine \eqref{app:eq:path-area} with
Corollary~\ref{app:cor:renyi}. The entropy-index-dependent choice in two
dimensions changes the aspect ratio, not the uniform path estimates.
In dimensions at least three no such choice is needed. For
$\alpha\ge1$, use $S_\alpha\le S_1$.
\end{appproof}

This is an explicit failure of a R\'enyi area law within the class
admitting these uniformly controlled gapped paths, not merely a failure
of one method to prove it. The paths do not imply exact preparation by
a constant-depth, strictly finite-range circuit. Nor do they impose
translation invariance, a size-independent lower bound on every
nonzero coupling, or stability of the R\'enyi volume law under arbitrary
perturbations.
\clearpage
\section{One-dimensional thermal lower bounds and Gibbs compression}
\label{app:thermal}

\subsection{Overview and finite-size statement}

We transfer the one-dimensional Schmidt-tail bound to sufficiently cold
Gibbs states and obtain lower bounds on their R\'enyi entanglement of
purification and on trace-norm compression. Throughout this appendix,
$H_m$ is the strength-normalized row Hamiltonian of
Appendix~\ref{app:row}, $\Omega_m$ its ground state, and
$\rho_{\beta,m}=e^{-\beta H_m}/\operatorname{tr}e^{-\beta H_m}$.
The cut is the central cut of the row. Define
\begin{align}
 E_{p,\alpha}(\rho):=
 \inf_{\substack{L',R',\,|\Psi\rangle\in LL'RR'\\
                 \operatorname{tr}_{L'R'}|\Psi\rangle\langle\Psi|=\rho}}
 S_\alpha(\Psi_{LL'}),\qquad
 \kappa_\alpha:=\frac{\alpha}{1-\alpha},\qquad 0<\alpha<1.
 \label{app:eq:thermal-Ep-def}
\end{align}
The infimum includes all finite ancillary dimensions and all ancillary
bipartitions~\cite{doi:10.1063/1.1498001,PhysRevX.11.011047}.
For a normalized pure state, write
$T_\Omega(R):=\inf_{\operatorname{SR}(v)\le R}\|\Omega-v\|^2
=\sum_{j>R}\lambda_j^2$, allowing unnormalized approximants $v$.

\begin{appthm}[One-dimensional finite-temperature transfer]
\label{app:thm:thermal-transfer}
There are constants $C,\Delta_0>0$, independent of $m,\beta,\alpha$,
such that, for every $m\ge3$ and $0<\alpha<1$,
\begin{align}
 \beta\Delta_0\ge32m\log6+Cm^{3/2}+\log16
 \quad\Longrightarrow\quad
 E_{p,\alpha}(\rho_{\beta,m})
 \ge m\log2-\log2-\kappa_\alpha(Cm^{3/2}+\log8).
 \label{app:eq:thermal-1d-finite}
\end{align}
The row has $n=32m$ sites, on-site dimension at most six, and physical
gap at least $\Delta_0$. Its local data and gap bound are uniform in
$m$. The lower bound applies to every purification in
\eqref{app:eq:thermal-Ep-def}.
\end{appthm}

\subsection{Near-pure transfer and the partition function}

For real $x$, write $(x)_+:=\max\{x,0\}$.
\begin{applemma}[Transfer of a Schmidt tail to every purification]
\label{app:lem:near-pure}
Let $\langle\Omega|\rho|\Omega\rangle=1-\delta>0$ and
$d_\delta:=\sqrt{2-2\sqrt{1-\delta}}\le\sqrt{2\delta}$.
For every integer $R\ge1$ and $0<\alpha<1$,
\begin{align}
 E_{p,\alpha}(\rho)\ge\log R+
 \kappa_\alpha\log\left(\sqrt{T_\Omega(R)}-d_\delta\right)_+^2.
 \label{app:eq:near-pure-transfer}
\end{align}
A zero argument of the logarithm gives no useful bound for that $R$.
\end{applemma}

\begin{appproof}
For any purification $\Psi\in LL'RR'$, define the normalized ancillary
vector $\eta$ by
$(\langle\Omega|\otimes I)|\Psi\rangle=\sqrt{1-\delta}\,|\eta\rangle$.
Then $\|\Psi-\Omega\otimes\eta\|=d_\delta$, and $\eta$ may itself be
entangled across $L'|R'$. If $p_i,q_j$ are the Schmidt probabilities of
$\Omega,\eta$, those of $\Omega\otimes\eta$ are $p_iq_j$. The sum of
any $R$ of these products is at most the sum of the $R$ largest $p_i$:
group the selected pairs by $i$ and use $\sum_jq_j=1$.
Hence $T_{\Omega\otimes\eta}(R)\ge T_\Omega(R)$. Distance to the set
of vectors of Schmidt rank at most $R$ is $1$-Lipschitz, giving
\begin{align}
 T_\Psi(R)\ge
       \left(\sqrt{T_\Omega(R)}-d_\delta\right)_+^2.
 \label{app:eq:purification-tail}
\end{align}
For Schmidt probabilities $s_1\ge s_2\ge\cdots$, put
$M_\alpha=\sum_js_j^\alpha$. Then
\begin{align}
 s_{R+1}\le(M_\alpha/R)^{1/\alpha},\qquad
 T(R)\le s_{R+1}^{1-\alpha}\sum_{j>R}s_j^\alpha
       \le M_\alpha^{1/\alpha}R^{-(1-\alpha)/\alpha}.
 \label{app:eq:tail-to-entropy}
\end{align}
Thus $S_\alpha\ge\log R+\kappa_\alpha\log T(R)$.
Apply this to \eqref{app:eq:purification-tail} and take the infimum
over all purifications and ancillary bipartitions.
\end{appproof}

\begin{appproof}[Proof of Theorem~\ref{app:thm:thermal-transfer}]
The full excited-state weight satisfies
\begin{align}
 \delta_\beta:=1-\langle\Omega_m|\rho_{\beta,m}|\Omega_m\rangle
 =\frac{\sum_{j\ge1}e^{-\beta(E_j-E_0)}}
        {1+\sum_{j\ge1}e^{-\beta(E_j-E_0)}}
 \le(6^{32m}-1)e^{-\beta\Delta_0}.
 \label{app:eq:thermal-excited-weight}
\end{align}
This counts all physical excitations, including legal, non-history and
illegal sectors. Set $F=Cm^{3/2}$. The cooling condition in
\eqref{app:eq:thermal-1d-finite} implies $\delta_\beta\le e^{-F}/16$.
The one-dimensional Bell-sector construction gives
$T_{\Omega_m}(2^{m-1})\ge\tfrac12e^{-F}$. Consequently,
\begin{align}
 \left(\sqrt{T_{\Omega_m}(2^{m-1})}-d_{\delta_\beta}\right)_+^2
 \ge\left(\sqrt{e^{-F}/2}-\sqrt{e^{-F}/8}\right)^2
 =e^{-F}/8.
 \label{app:eq:thermal-tail-constant}
\end{align}
Lemma~\ref{app:lem:near-pure} with $R=2^{m-1}$ proves the claim.
\end{appproof}

\subsection{Joint dependence on temperature and R\'enyi index}

\begin{appcor}[Joint one-dimensional thermal lower bound]
\label{app:cor:thermal-1d}
There exist constants $c>0$, $\beta_0>0$ and $0<\alpha_0<1$ such that,
for every $\beta\ge\beta_0$ and $0<\alpha\le\alpha_0$,
\begin{align}
 \sup_{m\ge3}E_{p,\alpha}(\rho_{\beta,m})
 \ge c\min\!\left\{\beta^{2/3},\left(\frac{1-\alpha}{\alpha}\right)^2\right\}.
 \label{app:eq:thermal-1d-joint}
\end{align}
For a sufficiently small fixed $a>0$, taking
$\alpha=a\beta^{-1/3}$ and a suitable $m=\Theta(\beta^{2/3})$ gives
\begin{align}
 E_{p,\alpha}(\rho_{\beta,m})=\Theta(\beta^{2/3}).
 \label{app:eq:thermal-1d-diagonal}
\end{align}
The system size is selected for the prescribed $(\beta,\alpha)$.
\end{appcor}

\begin{appproof}
Write $\kappa=\kappa_\alpha$ and
$R_0=\min\{\beta^{2/3},\kappa^{-2}\}$. Choose $0<\zeta<1$ such that
\begin{align}
 32\zeta\log6+C\zeta^{3/2}\le\Delta_0/2,
 \qquad C\sqrt\zeta\le\log2/4.
 \label{app:eq:thermal-1d-zeta}
\end{align}
Set $m=\lfloor\zeta R_0\rfloor$. Increase $\beta_0\ge1$ and decrease
$\alpha_0$ so that $\zeta R_0\ge32$, $\kappa\le1$, and
$\beta\Delta_0/2\ge\log16$ throughout the stated range.
Then $m\ge16$, $m\ge\zeta R_0/2$, and $m\le\zeta\beta^{2/3}$;
the first inequality in \eqref{app:eq:thermal-1d-zeta} gives the
cooling condition in \eqref{app:eq:thermal-1d-finite}.
Also $\kappa\sqrt m\le\sqrt\zeta$, so
$\kappa Cm^{3/2}\le m\log2/4$ and
$\log2+\kappa\log8\le4\log2\le m\log2/4$.
Theorem~\ref{app:thm:thermal-transfer} yields
$E_{p,\alpha}\ge\tfrac12m\log2\ge\tfrac14\zeta\log2\,R_0$.
For $\alpha=a\beta^{-1/3}$ with sufficiently small fixed $a$,
$R_0=\Theta(\beta^{2/3})$. Placing the entire purifying ancilla on the
right gives the dimension upper bound
$E_{p,\alpha}\le\log\dim\mathcal H_L=16m\log6$, proving
\eqref{app:eq:thermal-1d-diagonal} on the selected family.
\end{appproof}

\subsection{A Gibbs trace-norm compression lower bound}

\begin{appprop}[Joint temperature--accuracy lower bound in one dimension]
\label{app:prop:gibbs-compression}
There are constants $B,c>0$ and $L_0$ such that, for every $L\ge L_0$,
with $\epsilon=e^{-L}$ and $\beta=BL$, one can choose a row with
$m=\Theta(L^{2/3})$ for which every operator $X$ obeying
\begin{align}
 \|X-\rho_{\beta,m}\|_1\le\epsilon
 \quad\text{has}\quad
 \log\operatorname{SR}(X)\ge cL^{2/3}
 \label{app:eq:gibbs-rank}
\end{align}
across the central cut. The operator need not be positive, normalized,
Hermitian, or exactly ground-state preserving.
\end{appprop}

\begin{appproof}
Choose a fixed small $a>0$ with $Ca^{3/2}\le1$ and set
$m=\lfloor aL^{2/3}\rfloor\ge3$, so $F=Cm^{3/2}\le L$.
A fixed sufficiently large $B$ gives
$BL\Delta_0\ge L+32m\log6+\log2$ for every $L\ge L_0$.
By \eqref{app:eq:thermal-excited-weight} and commutation with
$P_0=|\Omega_m\rangle\langle\Omega_m|$,
\begin{align}
 \|\rho_{\beta,m}-P_0\|_1=2\delta_\beta\le e^{-L},
 \qquad \|X-P_0\|\le2e^{-L}.
 \label{app:eq:gibbs-ground-distance}
\end{align}
Choose the phase so $a_o=\langle\Omega_m|o\rangle\ge1/\sqrt2$.
Since $o$ is a product state, $v=Xo/a_o$ has
$\operatorname{SR}(v)\le\operatorname{SR}(X)$ and
$\|v-\Omega_m\|^2\le8e^{-2L}$.
If $\operatorname{SR}(X)\le2^{m-1}$, the established Schmidt-tail
bound gives instead
\begin{align}
 \|v-\Omega_m\|^2\ge\tfrac12e^{-Cm^{3/2}}
                      \ge\tfrac12e^{-L},
 \label{app:eq:gibbs-contradiction}
\end{align}
contradicting $L>\log16$. Thus $\operatorname{SR}(X)>2^{m-1}$.
\end{appproof}

The same lower bound applies to an MPO's bond dimension at this cut.
Along $\beta=\Theta(L)$ it also gives the necessary exponent conditions
\begin{align}
 \begin{aligned}
 \log\operatorname{SR}(X)\lesssim\beta^a+L^b
 &\quad\Longrightarrow\quad\max\{a,b\}\ge\tfrac23,\\
 \log\operatorname{SR}(X)\lesssim\beta^aL^b
 &\quad\Longrightarrow\quad a+b\ge\tfrac23,
 \end{aligned}
 \label{app:eq:gibbs-exponent-constraints}
\end{align}
for universal bounds of these forms, also allowing fixed
multiplicative polylogarithmic factors.

\end{document}